\documentclass[a4paper,11pt]{article}
\usepackage{jcappub, bm, color} 
\usepackage{amssymb,amsfonts,slashed,amsthm,amsmath,graphicx, soul}
\usepackage[caption=false]{subfig}
\title{CMB spectral distortions from \\ black holes formed by vacuum bubbles}

\author{Heling Deng,}
\author{Alexander Vilenkin,}
\author{and Masaki Yamada}

\affiliation{Institute of Cosmology, Department of Physics and Astronomy, 
Tufts University, Medford, MA  02155, USA}

\emailAdd{Heling.Deng@tufts.edu}
\emailAdd{vilenkin@cosmos.phy.tufts.edu}
\emailAdd{Masaki.Yamada@tufts.edu}

\newcommand{\beq}{\begin{equation}}
\newcommand{\eeq}{\end{equation}}

\def\({\left(}
\def\){\right)}
\def\[{\left[}
\def\]{\right]}

\def\lmk{\left(}
\def\rmk{\right)}

\newcommand{\eq}[1]{Eq.~(\ref{#1})}
\newcommand{\bel}[1] {\begin{equation}\label{#1}}
\newcommand{\beal}[1] {\begin{eqnarray}\label{#1}}
\newcommand{\be}{\begin{equation}}
\newcommand{\ee}{\end{equation}}
\newcommand{\bea}{\begin{eqnarray}} 
\newcommand{\eea}{\end{eqnarray}}
\newcommand{\abs}[1]{\left\vert#1\right\vert}

\abstract{
Vacuum bubbles may nucleate and expand during the cosmic inflation.  When inflation ends, the bubbles run into the ambient plasma, producing strong shocks followed by underdensity waves, which propagate outwards.  The bubbles themselves eventually form black holes with a wide distribution of masses.  It has been recently suggested that such black holes may account for LIGO observations and may provide seeds for supermassive black holes observed at galactic centers.  They may also provide a significant part or even the whole of the dark matter.  We estimate the spectral $\mu$-distortion of the CMB induced by expanding shocks and underdensities. The predicted distortions averaged over the sky are well below the current bounds, but localized peaks due to the largest black holes impose constraints on the model parameters.
}

\begin{document}
\maketitle
\flushbottom

\section{Introduction}

Supermassive black holes (SMBH) reside at the centers of most large galaxies, as well as in some dwarf galaxies~\cite{LyndenBell:1969yx, Kormendy:1995er}.  Their masses range from $\sim 10^6 M_\odot$ to $\sim 10^{10}M_\odot$, and observations of quasars indicate that many of them are already in place at redshifts as high as $z\sim 6-7$ (See Ref. \cite{Kormendy:2013dxa} for a review.)  The origin of SMBH is rather puzzling:  with standard assumptions about accretion onto black holes, the available cosmic time is insufficient for SMBH to grow from stellar-mass seeds~\cite{Haiman:2004ve}.  One is therefore led to consider more exotic possibilities, such as nonstandard accretion models~\cite{Volonteri:2005fj, Alexander:2014noa, Pacucci:2015rwa, Inayoshi:2015pox} or a primordial origin of SMBH~\cite{Rubin:2001yw, Bean:2002kx, Duechting:2004dk, Clesse:2015wea, Carr:2018rid}. Here we focus on the second possibility, that is, the seeds for SMBH are primordial black holes (PBHs) formed in the early Universe.

Another motivation to consider PBHs comes from recent LIGO observations \cite{Bird:2016dcv, Clesse:2016vqa, Sasaki:2016jop, Eroshenko:2016hmn, Carr:2016drx}.  LIGO detected gravitational waves emitted by inspiraling and merging black holes of mass $\sim 30 M_\odot$ \cite{Abbott:2016blz, Abbott:2016nmj, Abbott:2017vtc, Abbott:2017oio, Abbott:2017gyy}.  This mass is somewhat larger than expected, but more surprisingly, these black holes appear to have rather low spins, with all but one being consistent with zero spin.  This is in contrast with current observations and theoretical expectations, both favoring high spin values for black holes formed by stellar collapse \cite{Miller:2011jq, Gammie:2003qi}. On the other hand, PBHs may have relatively low spins in some production mechanisms.

It has also been suggested that dark matter observed in galaxies and clusters may be made up of primordial black holes.  This possibility is excluded for most values of the black hole mass, but some observationally allowed windows still remain~\cite{Carr:2016drx, Inomata:2017vxo} (see Sec.~III). 

PBHs can be formed by a variety of mechanisms; for a review and references see, e.g., \cite{Carr:2005zd}.  The most widely discussed scenario is PBH formation during the radiation era, induced by inflationary density perturbations~\cite{Ivanov:1994pa, GarciaBellido:1996qt, Kawasaki:1997ju, Yokoyama:1998pt, Garcia-Bellido:2017mdw, Hertzberg:2017dkh}. The Jeans length in the radiation era is comparable to the horizon, so black holes formed at cosmic time $t$ have mass $M\sim t$ (in units where $c=G=1$).  The density fluctuation required for a horizon-size region to collapse to a black hole is $\delta\equiv\delta\rho/\rho\sim 1$.
Even if PBH are formed only at high peaks of the density field, the rms fluctuation on the corresponding scale should be rather large, $\delta_{\rm rms}\gtrsim 0.1$ (assuming the fluctuations are Gaussian, as is usually the case in inflationary models).  One therefore has to assume that the inflaton potential has some feature that generates large fluctuations on a relatively narrow range of scales, while $\delta_{\rm rms}\sim 10^{-5}$ on scales accessible to CMB and large-scale structure observations.  However, even with this restriction the PBH model of SMBH formation may have a serious problem. 

Density fluctuations in the radiation era oscillate as sound waves and are dissipated by Silk damping on scales smaller than the photon diffusion length \cite{Silk:1967kq}.  Fluctuations dissipated during the redshift interval $5\times 10^4 < z < 2\times 10^6$ are not completely thermalized and generate so-called $\mu$-distortions in the CMB spectrum~\cite{Sunyaev:1970er, Daly, Barrow_1991, Hu:1992dc, Chluba:2012we}.  The comoving wavelengths of such fluctuations correspond to PBH masses $\sim 10^5 - 10^9 M_\odot$, and it was shown in Refs.~\cite{Kohri:2014lza, Nakama:2017xvq} that strong observational bounds on $\mu$-distortions rule out the formation of such PBH in any appreciable numbers.  

This problem may be circumvented by considering models with strongly non-Gaussian fluctuations.  For example, Nakama {\it et al.} proposed a scenario where large density fluctuations are generated only in isolated patches, comprising a very small fraction of the total volume \cite{Nakama:2016kfq}.  However, inflationary models implementing this scenario tend to be rather contrived.
Another possibility is to have PBH formed with $M < 10^5 M_\odot$ and then grow by accretion.  However, the seeds for SMBH have to be at least as massive as $\sim 10^3 - 10^4 M_\odot$ \cite{Duechting:2004dk, Tanaka:2008bv}.  On the other hand, inflationary models typically yield a mass distribution of PBH extending over several orders of magnitude, and it may be hard to find a model that would have a large density of PBH with $M \sim 10^4 M_\odot$ and a negligible density at $M\sim 10^5 M_\odot$.

In this paper we shall discuss the model developed in Refs.~\cite{Garriga:2015fdk, Deng:2017uwc}, where vacuum bubbles nucleate during inflation and later form PBH after inflation ends.\footnote{
There are some other scenarios to have PBH formed: cosmic strings~\cite{Hawking:1987bn, Garriga:1992nm, Caldwell:1995fu} bubble collisions~\cite{Hawking:1982ga}, and domain walls~\cite{Garriga:1992nm, Khlopov:2008qy, Garriga:2015fdk, Deng:2016vzb}. 
}
This model has several attractive features: {\it (i)} It can be naturally implemented in landscape models of the kind suggested by string theory \cite{VY}; {\it (ii)} It predicts a distinctive PBH mass spectrum, ranging over many orders of magnitude and depending on only two free parameters; {\it (iii)} With some parameter choices, it can account for SMBH, for the black hole mergers observed by LIGO, and/or for the dark matter; {\it (iv)} Black holes formed by this mechanism have zero spin; {\it (v)} Finally, it can satisfy the constraints imposed by the $\mu$-distortion observational bounds -- as we will show in this paper.

The reason the formation of PBH from Gaussian fluctuations is subject to $\mu$-distortion bounds is that ${\cal O}(1)$ density fluctuations at the sites of PBH formation imply a relatively large value of $\delta_{\rm rms}$ in the rest of space.  In our model, bubbles of a high-energy vacuum nucleate and expand during inflation, reaching ultra-relativistic expansion speeds.  When inflation ends, the expanding bubble walls run into the ambient radiation, and most of the kinetic energy of the walls is transferred to the radiation, producing powerful shock waves.  Space outside the shocks is unperturbed, but the energy carried by the shocks may be rather large, and may potentially result in a detectable $\mu$-distortion.  Estimating the magnitude of this distortion is our goal in the present paper.

The paper is organized as follows.  The model of PBH formation from bubbles nucleating during inflation is reviewed in Sec.~II and observational bounds on the model parameters are discussed in Sec.~III.  Propagation of shock waves produced by the bubbles is studied both analytically and numerically in Sec.~IV, and the resulting spectral distortions are estimated in Sec.~V.  Our conclusions are summarized in Sec.~VI.  Throughout this paper, we use the Planck units ($G = 1$).

\section{The scenario}

Models of inflation typically involve a scalar field $\phi$ -- the inflaton -- which slowly rolls down the slope of its potential $U(\phi)$, while $U(\phi)$ remains nearly constant and drives the inflationary expansion:
\beq
U(\phi)\approx \rho_i = {\rm const}.
\eeq
The field eventually ends up at the minimum of $U(\phi)$ corresponding to our vacuum.  In addition to the inflaton $\phi$, the underlying particle physics generally includes some other scalar fields.  We know that there should be at least one such field -- the Higgs field of the Standard Model.  Grand unified theories include a number of Higgs-like fields, and string theory suggests the existence of hundreds of scalar fields.  The inflaton then ``rolls'' in a multi-dimensional potential, including a number of minima in addition to our vacuum.  Bubble nucleation in this setting occurs during inflation if some of these minima have vacuum energy density $\rho_b$ lower than $\rho_i$ \cite{Coleman:1980aw}.\footnote{Nucleation of bubbles with $\rho_b >\rho_i$ is also possible, but it is strongly suppressed~\cite{Lee:1987qc}.}  We will be interested in the case where $\rho_b >0$.  (Bubbles with $\rho_b <0$, if they were formed in our universe, would be catastrophic: they would expand without bound, until they engulf the entire observable region.)  To simplify the analysis, we shall assume a separation of scales, $\rho_b\ll\rho_i$, but our conclusions will apply to the case of $\rho_b\sim\rho_i$ as well.  

Bubbles of lower-energy vacuum nucleate having a microscopic size and immediately start to expand. The difference in vacuum tension on the two sides of the bubble wall results in a force $F=\rho_i -\rho_b$ per unit area of the wall, so the bubble expands with acceleration, acquiring a large Lorentz factor.  However, the exterior region remains completely unaffected by the expanding bubble.  Bubbles formed at earlier times during inflation expand to a larger size,  so at the end of inflation the bubbles have a wide distribution of sizes \cite{Garriga:2015fdk},
\beq
n(R_i) = \Gamma {R_i^{-3}}.
\label{dn}
\eeq
Here, $n(R_i) = R_i dn/dR_i$ is the number density of bubbles of radius $\sim R_i$ and $\Gamma$ is the dimensionless bubble nucleation rate per Hubble volume per Hubble time. The distribution (\ref{dn}) has an effective lower cutoff \cite{Deng:2017uwc} at $R_{\rm min}\sim H_i^{-1}$, where $H_i = (8\pi \rho_i/3)^{1/2}$ is the expansion rate during inflation.  Note that large bubbles have radii much greater than the horizon, $R_i\gg H_i^{-1}$. We assume that $\Gamma$ is constant during inflation, which is usually the case for small-field inflation~\cite{VY}.  We shall comment on the possibility of a variable $\Gamma$ in Section V.

When inflation ends, the inflaton energy outside the bubble thermalizes into hot radiation.  Following Refs.~\cite{Garriga:2015fdk,Deng:2017uwc}, we shall assume for simplicity that thermalization occurs instantaneously at time $t_i$; then $H_i=1/2t_i$ and $\rho_i = 3/32\pi t_i^2$.  The energy of the bubble of radius $R_i$ immediately prior to thermalization is equal to the inflaton energy in the region displaced by the bubble,
\beq
E_i =\frac{4\pi}{3} \rho_i R_i^3 = \frac{1}{2} H_i^2R_i^3.
\eeq
For large bubbles, most of this energy is the kinetic energy of the rapidly expanding bubble wall.

The wall runs into the radiation and quickly loses much of its energy, producing a shock wave that propagates outwards.\footnote{As in Refs.~\cite{Garriga:2015fdk,Deng:2017uwc}, we assume that radiation is completely reflected from the wall.  If reflection is not perfect, the shock wave and the resulting spectral distortion would be weaker.  Hence our results can be regarded as an upper bound on spectral distortion.}  The wall then comes to rest with respect to the Hubble flow, expanding with a Lorentz factor $\gamma \approx H_i R_i$ (for $R_i\gg t_i$).  The remaining energy of the bubble is
\beq
E_b \approx \left( \frac{4\pi}{3}\rho_b + 4\pi \sigma H_i \right) R_i^3 \equiv \kappa R_i^3 ,
\label{Eb}
\eeq
where $\sigma$ is the wall tension (or mass per unit area).  The second term in the parenthesis comes from the energy of the wall (= $4 \pi \sigma \gamma R_i^2$). The energy difference $E_i - E_b$ is transferred to the shock wave.  

The bubble can now form a black hole by one of the two different scenarios, depending on the bubble size.  (i) The bubble wall is pulled inwards by the interior vacuum tension, the wall tension, as well as the exterior radiation pressure; so it expands to some maximal radius, then shrinks and eventually collapses to a Schwarzschild singularity.  Such bubbles are called subcritical.  (ii) If in the course of bubble expansion its radius exceeds the interior de Sitter horizon $H_b^{-1}=(8\pi \rho_b/3)^{-1/2}$, the bubble interior begins to inflate.  The bubble then continues to expand without bound, and a wormhole is created connecting the inflating bubble universe with the radiation-dominated FRW parent universe.  We shall refer to such bubbles as supercritical.  In either case, the resulting central object is seen as a black hole by an FRW observer.  

For relatively small bubbles, we can expect the black hole mass to be $M\approx E_b$.  The reason is that some layer in the bubble exterior is initially evacuated by the impact.  The bubble wall then accelerates inward, away from the radiation, so there is essentially no contact of the bubble with radiation all the way until the black hole is formed.  The work done by radiation on the bubble can therefore be neglected and the energy (\ref{Eb}) is conserved.\footnote{A conserved energy in bubble spacetimes was defined in Ref.~\cite{Berezin:1982ur} and was extensively used in \cite{Garriga:2015fdk,Deng:2017uwc}.}  However, it was noted in Ref.~\cite{Garriga:2015fdk} that the dependence $M\propto R_i^3$ cannot extend to arbitrarily large $R_i$.
The region affected by the bubble (enclosed by the shock front) comes within the horizon at $t_H \sim a (t_H) R_i$ or $t_H \sim H_i R_i^2$ during the radiation dominated epoch, where $a(t) \propto \sqrt{t}$.  At that time the black hole should already be in place, and thus its mass is bounded by the energy enclosed by the Hubble volume, 
\beq
M\lesssim 
\frac{4\pi}{3} \rho (t_H) H^{-3} (t_H) \sim 
H_i R_i^2 .
\label{bound}
\eeq
The estimate $M\sim \kappa R_i^3$ cannot therefore apply for $R_i\gg H_i /\kappa$ or $M\gg M_*$, where  
\beq
M_* \sim H_i^3 /\kappa^2.
\eeq

Numerical simulations in Ref.~\cite{Deng:2017uwc} show that the black hole mass is indeed well approximated by $E_b$ for small $M$ and that the bound (\ref{bound}) is saturated for large $M$.  Hence, as a rough estimate we can use
\beq
M \sim \kappa R_i^3 ~~~ {\rm for} ~M\lesssim M_*      
\label{M1}
\eeq
and
\beq
M\sim H_i R_i^2  ~~~ {\rm for} ~M\gtrsim M_* . 
\label{M2}
\eeq
In the latter case, the black hole is formed at the time $t_H\sim H_i R_i^2$, with its Schwarzschild radius comparable to the horizon.  It can be shown that black holes with $M>M_*$ are necessarily supercritical and have inflating universes inside.

The mass distribution of black holes is conveniently characterized by the quantity
\beq
f(M)=\frac{M^2}{\rho_{\rm CDM}}\frac{dn}{dM} ,
\eeq
which gives the fraction of dark matter density $\rho_{\rm CDM}$ in black holes of mass $\sim M$.  Since the black hole density and $\rho_{\rm CDM}$ are diluted in the same way, $f(M)$ remains constant in time. We note that during the radiation era ($t<t_{\rm eq}$) the dark matter density is of the order
\beq
\rho_{\rm CDM}(t) \sim (Bt^{3/2}M_{\rm eq}^{1/2})^{-1}, 
\eeq
where $B\sim 10$ is a numerical coefficient and $M_{\rm eq}\sim 10^{17} M_\odot$ is the dark matter mass within the horizon at $t_{\rm eq}$. Then from Eqs.~(\ref{dn}), (\ref{M1}), and (\ref{M2}), $f(M)$ is given by \cite{Deng:2017uwc}
\beq
f(M)\sim B\Gamma \left(M_{\rm eq}/M\right)^{1/2}
\label{fM1}
\eeq
for $M>M_*$ and
\beq
f(M)\sim B\Gamma \left(M_{\rm eq}/M_* \right)^{1/2}
\label{fM2}
\eeq
for $M<M_*$.  The distribution (\ref{fM2}) has an effective lower cutoff at
\beq
M_{\rm min} \sim \kappa R_{\rm min}^3 \sim \kappa H_i^{-3}.
\eeq
Depending on the microphysics, the characteristic masses $M_*$ and $M_{\rm min}$ can take a wide range of values.

\section{Observational constraints}

\begin{figure}
   \centering
\includegraphics[scale=0.35]{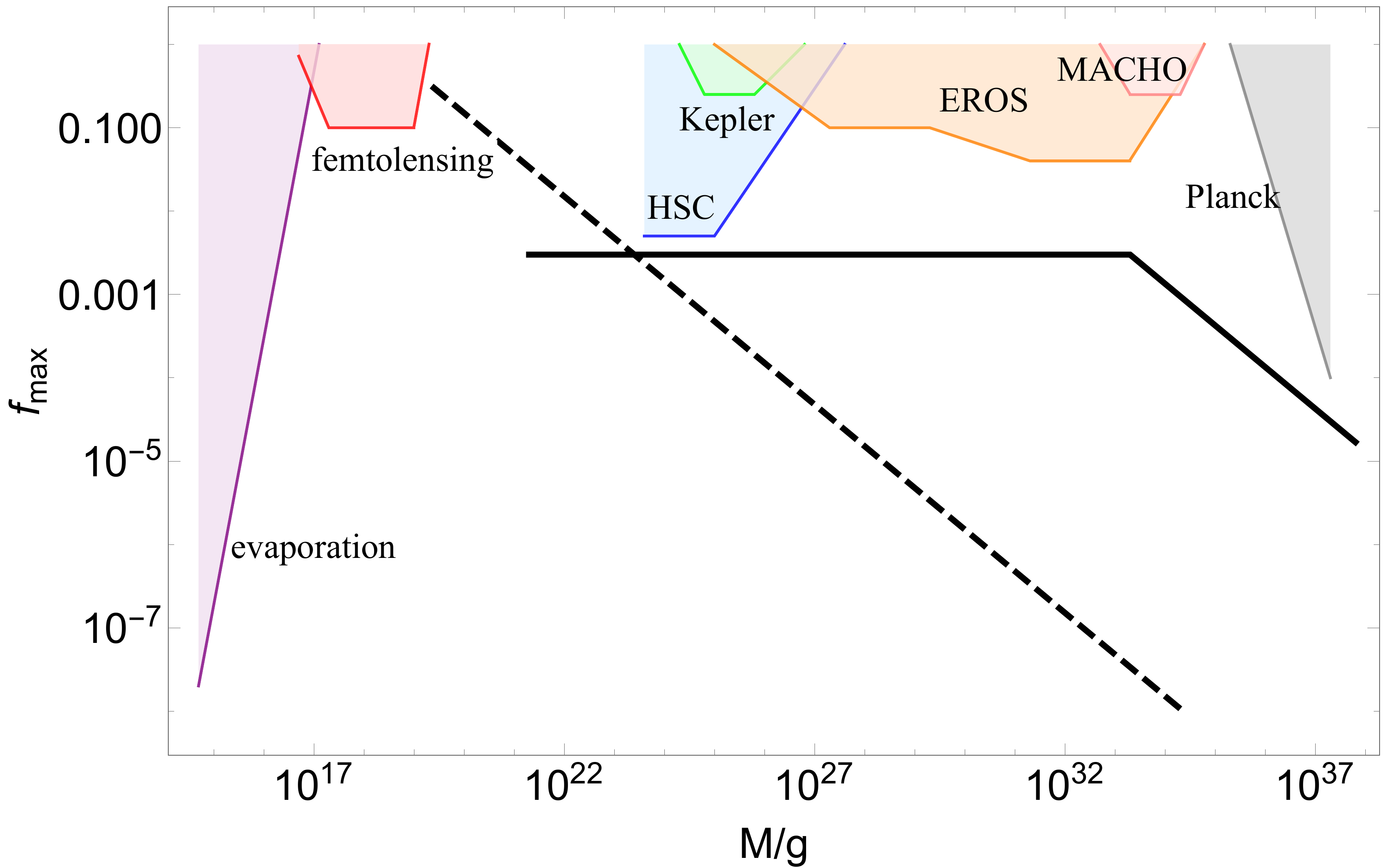}
\caption{\label{PBH}{A sketch of constraints from different observations on the fraction of dark matter in PBHs as a function of the PBH mass for a monochromatic mass distribution. More detail can be found in e.g. Ref. \cite{Carr:2016drx,Carr:2017jsz,Inomata:2017vxo} and references therein. As an illustration, we also show the PBH distributions $f(M)$ for our model with $\Gamma= 10^{-12}$, $M_* = M_\odot$ and $M_{\rm min} = 10^{-12}M_\odot$ (solid line), and with $\Gamma= 10^{-17}$, $M_*= M_{\rm min} = 10^{19}$ g (dashed line).}}
\end{figure}

Observational bounds on $f(M)$ in different mass ranges have been extensively studied (see, e.g., Ref.~\cite{Carr:2016drx} for a recent review).  The current bounds are summarized in Fig.~\ref{PBH}.  As an illustration, we also show distributions of the form (\ref{fM1}), (\ref{fM2}) with two different sets of parameters. Strictly speaking, the bounds in Fig.~\ref{PBH} apply to a "monochromatic" black hole distribution, where all black holes have a similar mass.  Bounds for extended distributions were discussed in \cite{Kuhnel:2017pwq, Carr:2017jsz} and were applied to our model in \cite{Deng:2017uwc}, but the difference is not significant by order of magnitude.

In order to have PBH merger rate suggested by LIGO observations, we need \cite{Sasaki:2016jop}
\beq
f(M\sim 30 M_\odot)\sim 10^{-3}.
\eeq
This condition can be consistent with the Planck satellite constraint~\cite{Ricotti:2007au, Ali-Haimoud:2016mbv} only if 
\beq
M_* \lesssim 10^2 M_\odot, ~~~ \Gamma \sim 10^{-12}.
\label{Planck satellite constraint}
\eeq

The number density of PBH of mass $\sim M>M_*$ at the present time is
\beq
n_M\sim B\Gamma\left(\frac{M_{\rm eq}}{M}\right)^{1/2}\frac{\rho_{\rm CDM}}{M} \sim 10^{20} \Gamma \left(\frac{M}{M_\odot} \right)^{-3/2} {\rm Mpc}^{-3},
\label{nM}
\eeq
where $\rho_{\rm CDM}$ is the dark matter density today. The seeds for SMBH should have density $n_M \sim 0.1 \, {\rm Mpc}^{-3}$.  From Eq.~(\ref{nM}), the mass of such black holes in our scenario is $M_{\rm seed}\sim 10^{14} \Gamma^{2/3} M_\odot$ (assuming that $M_{\rm seed}>M_*$).  Requiring that $M_{\rm seed}\gtrsim 10^3 M_\odot$, we need $\Gamma\gtrsim 10^{-17}$. The largest black hole we can expect to find in our Hubble region (of radius $10^4$~Mpc) has mass $M \sim 10^{21} \Gamma^{2/3} M_\odot$.

Another important condition comes from the Hawking radiation constraint~\cite{Carr:2009jm}.  In order to have a non-negligible black hole density and avoid this constraint, we should require that $M_{\rm min} > 10^{15}$ g.  This yields the condition
\beq
\eta_b^2 \gtrsim 10^{10} \eta_i^3/M_p,
\label{ineq}
\eeq
where $M_p\sim 10^{19}$~GeV is the Planck mass and we have defined the microphysics energy scales $\eta_i$ and $\eta_b$ according to $\rho_i=\eta_i^4$, $\rho_b = \eta_b^4$.  We have also assumed for simplicity that the second term in the parentheses of Eq.~(\ref{Eb}) is negligible, so $\kappa\sim \rho_b$.  With $\eta_b<\eta_i$, Eq.~(\ref{ineq}) implies $\eta_i \lesssim 10^9$ GeV.  Hence our scenario requires a relatively low energy scale of inflation.

We note also that it follows from the expressions for $M_*$ and $M_{\rm min}$ that 
\beq
\frac{M_*}{M_{\rm min}}\sim \left(\frac{\eta_i}{\eta_b}\right)^{12}.
\eeq
With $M_*\lesssim 10^2 M_\odot$, $M_{\rm min}\gtrsim 10^{15}$ g, this gives $\eta_i/\eta_b \lesssim 10^2$.  An example of microphysics parameters that satisfy observational constraints and may account for SMBH and LIGO data is $\eta_i \sim 10^4$ GeV, $\eta_b \sim 10^3$ GeV, $\Gamma\sim 10^{-12}$.  In this case, $M_*\sim M_\odot$, $M_{\rm min}\sim 10^{21}$ g, $f(M\sim 30 M_\odot)\sim 10^{-3}$, and $M_{\rm seed}\sim 10^6 M_\odot$.  The corresponding mass distribution is shown by a solid line in Fig.~\ref{PBH}.  Note that any set of parameters that can account for LIGO results would also provide sufficiently massive seeds for SMBH.

\begin{figure}
   \centering
\includegraphics[scale=1]{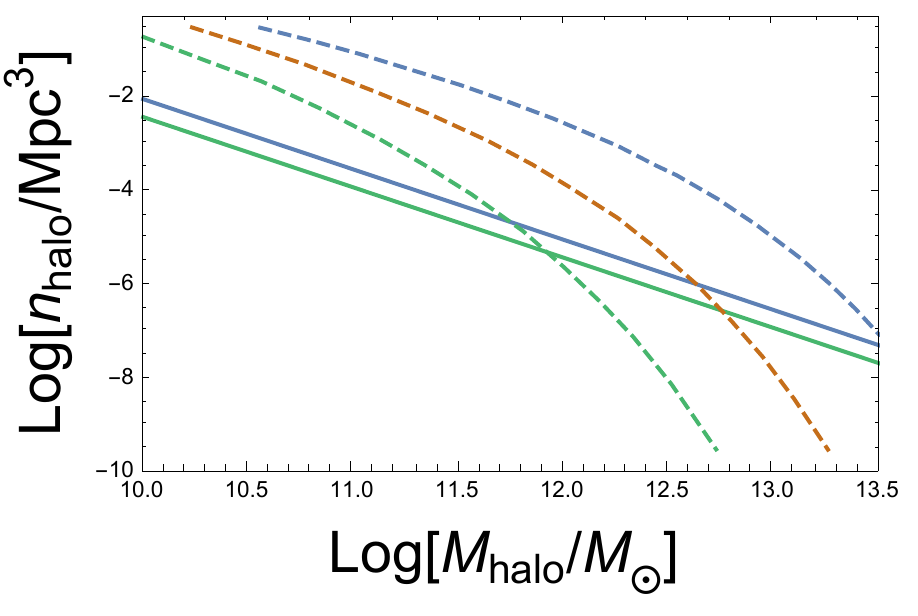}
\caption{\label{halo}
Halo number densities (\ref{nhalo}) with $\Gamma=10^{-12}$ at $z = 4, 8$ (solid lines).  The Sheth-Tormen halo distributions for $z = 4 ,6, 8$ are shown by dashed lines. 
In both cases the redshift increases from top to bottom.  
}
\end{figure}

The black hole distribution (\ref{nM}) may have a significant effect on structure formation in the universe.  Once a black hole is formed, it attracts a dark matter halo from its surroundings.  The halo does not grow much during the radiation era and grows like $(1+z)^{-1}$ during the matter era, so the halo mass at redshift $z<z_{\rm eq}$ is~\cite{Mack_2007} $M_{\rm halo} \sim M (1+z_{\rm eq})/(1+z)$.  Here, $z_{\rm eq}\sim 4000$ is the redshift at the time of equal matter and radiation densities.  The halo mass distribution is then given by 
\beq
n_{\rm halo} (M_{\rm halo}, z) \sim 10^{11} \Gamma \ {\rm Mpc}^{-3} \lmk \frac{M_{\rm halo} }{ 10^{10} M_\odot} \rmk^{-3/2} (1+z)^{-3/2}, 
\label{nhalo}
\eeq
This distribution with $\Gamma\sim 10^{-12}$ is plotted at several redshifts in Fig.~\ref{halo}, together with the Sheth-Tormen distribution \cite{Sheth_1999} describing the halo mass function in the standard hierarchical structure formation scenario.   The black hole halos are subdominant at small masses, but our model predicts early formation of very massive halos with $M_{\rm halo}\gtrsim 10^{12}M_\odot$ at $z\gtrsim 5$.  There are in fact some observational indications that the standard model does not account for the most massive halos at high redshifts (see, e.g., \cite{Kang:2015xhf, Steinhardt:2015lqa, Franck:2016aa}.)

Another interesting set of parameters is $\Gamma\sim 10^{-17}$ and $\eta_i \sim \eta_b \sim 10^7$ GeV; then $M_*\sim M_{\rm min} \sim 10^{19}$ g.  In this case $f(M_*)\sim 1$, so PBH may play the role of dark matter.  At the same time, $M_{\rm seed}\sim 10^3 M_\odot$, so they may also serve as seeds for SMBH.\footnote{Bounds imposed by HSC observations \cite{Niikura:2017zjd} appeared to rule out PBH dark matter with $M\sim 10^{20} - 10^{23}$~g.  However, these bounds have been recently revised (see Ref.~\cite{Inomata:2017vxo} and references therein), and this mass window now appears to be open.}  The mass distribution in this case is shown by a dashed line in Fig.~\ref{PBH}.

Not included in Fig.~\ref{PBH} are observational bounds due to $\mu$-distortion of the CMB spectrum.\footnote{Another strong but highly model-dependent constraint comes from the annihilation signals of dark matter in ultracompact minihalos (UCMHs) around PBHs~\cite{Lacki:2010zf}. If the main component of dark matter is a weakly interacting massive particle, the annihilation in UCMHs may produce highly luminous gamma-ray sources. The constraints on these fluxes give strong upper bounds on the abundance of PBHs. But since these constraints depend on the details of dark matter models, including the mass and annihilation cross section of dark matter, we disregard them in this paper. 
}
These bounds depend on the model of black hole formation.  For PBH formed by Gaussian density fluctuations, it was shown in Refs.~\cite{Kohri:2014lza, Nakama:2017xvq} that their number density in the mass range $10^5 - 10^9 M_\odot$ must be much lower than that required to seed SMBH.  In the rest of this paper we shall discuss the CMB spectral distortions induced by black holes formed by vacuum bubbles.

\section{Shock propagation}

\begin{figure}[!ht]
   \centering
\subfloat[]{\includegraphics[scale=0.27]{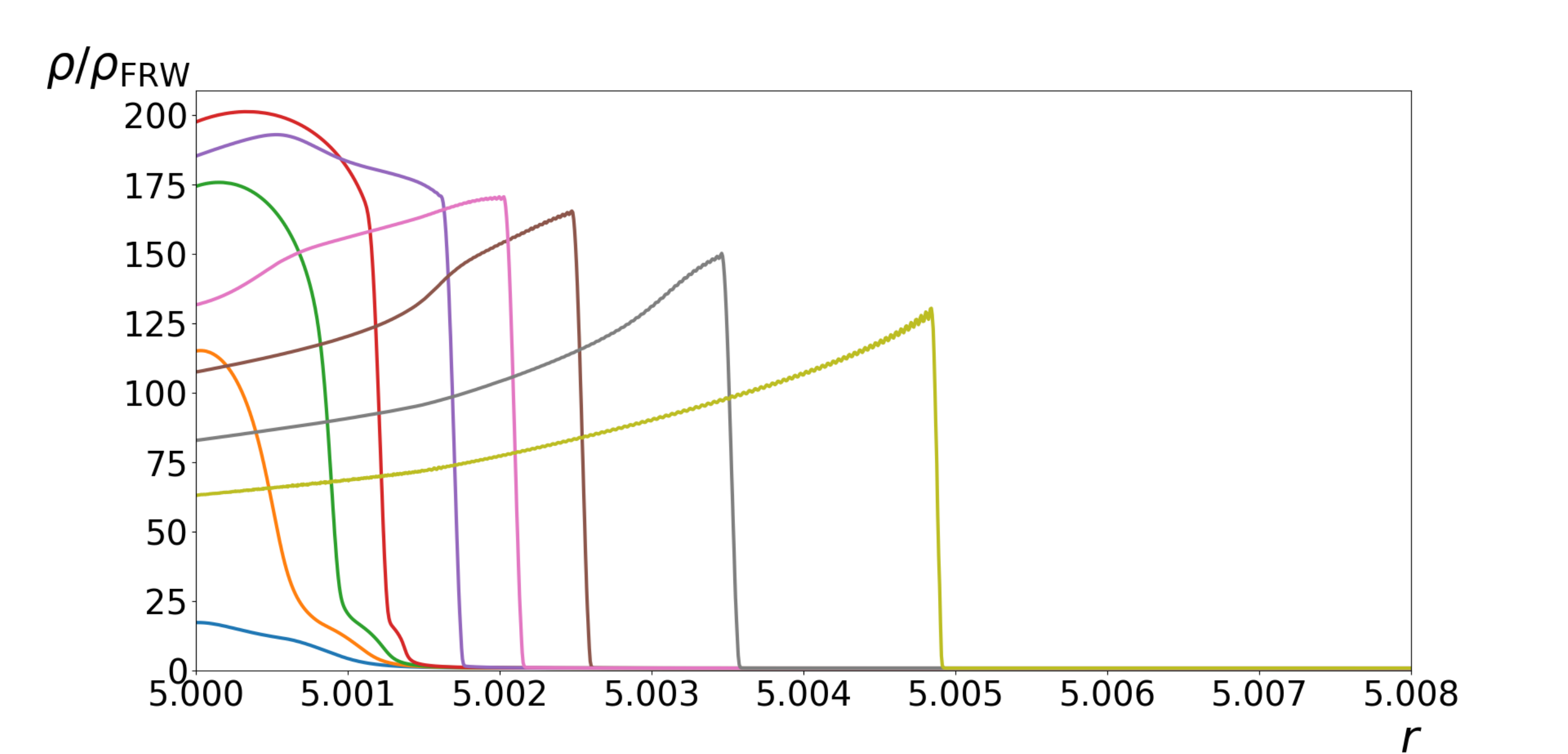}
}\subfloat[]{\includegraphics[scale=0.12]{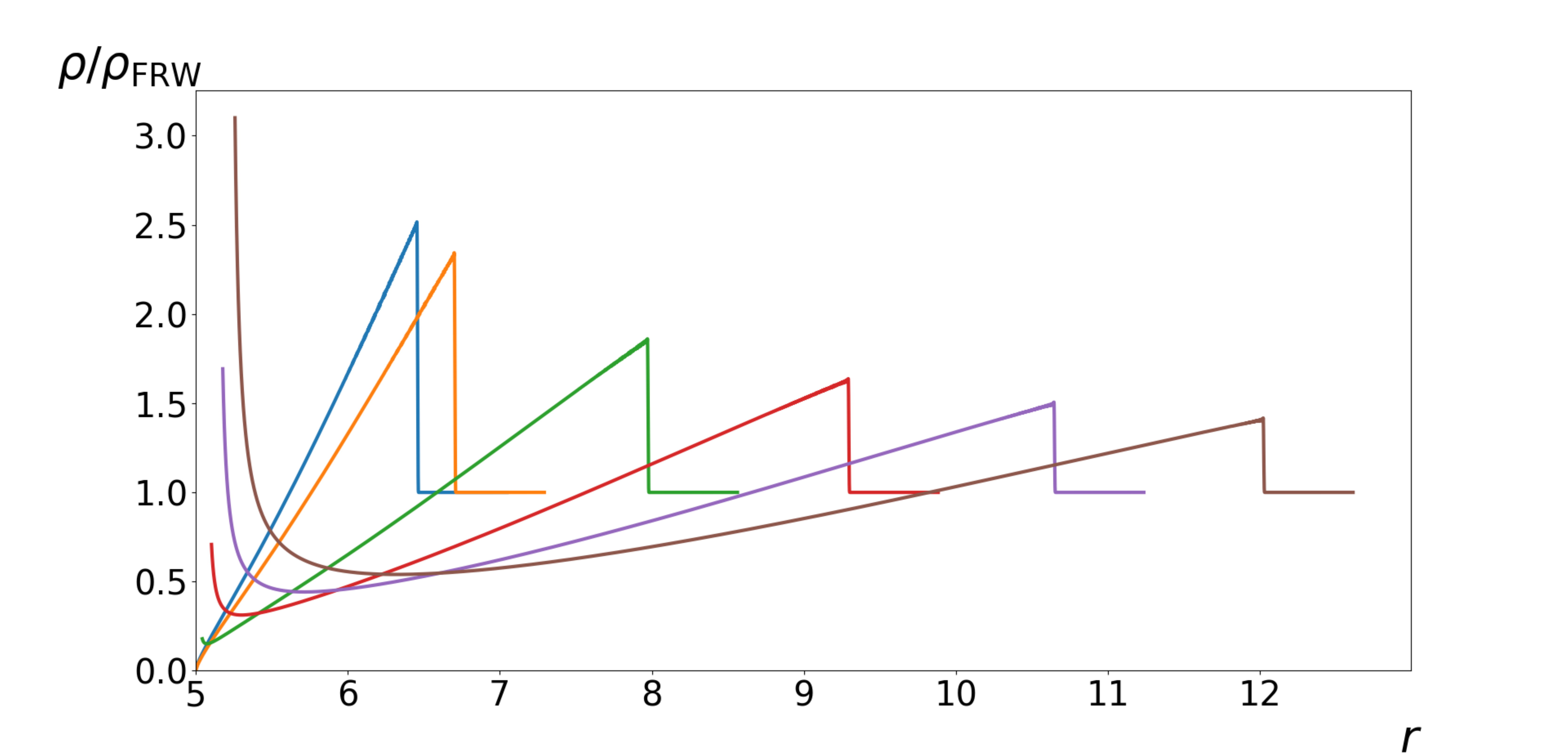}
}
\\
\subfloat[]{\includegraphics[scale=0.27]{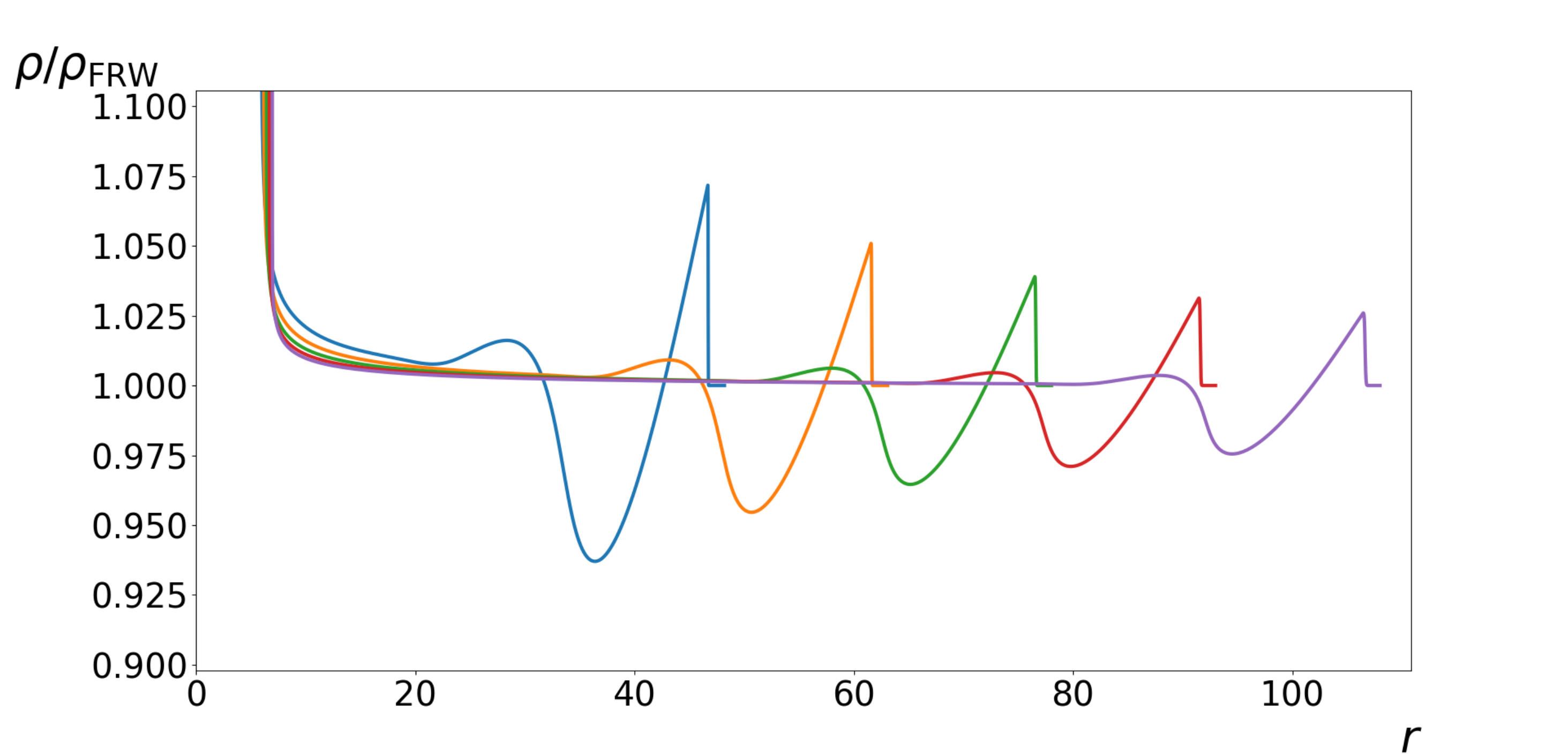}
}
\caption{\label{fig:The-radiation-energy}The radiation energy density $\rho$ as a function of the comoving radius $r$ at different moments of time outside of a subcritical bubble with $H_b=0.05H_{i}$, $\sigma\approx 0.005 H_i M_p^2$, and $R_{i}=5H_{i}^{-1}$. For all moments, $\rho$ has been rescaled so that the FRW density is 1. More details are given in the text. (a) $t\ll H_i^{-1}$. As the bubble hits the ambient fluid, an overdense shell forms and then propagates outwards in the form of a shock wave. Afterwards the density contrast across the shock begins to decline and the value of $\rho$ decreases rapidly close to the bubble wall. (b) $t \sim t_M$. The region around the bubble is evacuated by the impact.  The second (orange) profile corresponds to the time of black hole formation ($t = t_M$).  At that time, the value of $\rho$ on the apparent horizon is negligible compared to the FRW density (the black hole region is excised in simulations in order to avoid code crash). After black hole formation the evacuated region begins to fill in and the density at the apparent horizon grows to a large value. (c) $t\gg t_M$. An underdense wave propagates outwards following the shock.  At late times the overdense and underdense shells become small perturbations to the FRW density, and the region between the black hole and the shells comes back to FRW density.}
\end{figure}

We used the simulation code developed in Ref.~\cite{Deng:2017uwc} to study the evolution of the shocked region around the bubble nucleation site.   Several snapshots of a typical simulation for a subcritical bubble are shown in Fig. \ref{fig:The-radiation-energy}.\footnote{We will be mostly interested in supercritical bubbles, but the qualitative features of shock propagation are similar in both cases.  We used a subcritical bubble here because we could follow the shock evolution to a later time.}  More details will be given in Subsection D.  

As the bubble wall hits ambient radiation, almost all the wall energy is transferred to a thin radiation shell right outside the wall.  It follows from the bound (\ref{bound}) that for large bubbles, $R_i\gg t_i$, this energy is much greater than the mass of the black hole that will be left behind, $E_i \gg M$.  The energy transfer occurs on a very short time scale, $\Delta t\ll t_i$.\footnote{This may be an artifact of our assumption that thermalization occurs instantaneously.  In a more realistic model, we might have $\Delta t\sim t_i$, but the density in the shell would still be much higher than $\rho_r$.  Our results are not sensitive to the assumption about the thickness of the shell.} 
Hence the thickness of the shell is very small, $s_{+i} \sim \Delta t$ and its density is very high,
\beq
\rho_{+i} \sim \frac{R_i}{3s_{+i}}\rho_i \gg \rho_i.
\label{rhosi}
\eeq
We use the subscript "+" in reference to the overdense shell and subscript "-" in reference to the underdense shell that will be discussed below. The overdense shell develops a sharp shock front, which then propagates outwards.  

The radiation density in the region enclosed by the expanding shell is initially significantly below the density $\rho_r$ in the unperturbed FRW region.  This is due to the impact of the bubble wall and to its subsequent inward acceleration.  This region begins to fill up at the time $t_M\sim M$, when the black hole apparent horizon is formed. The result is that an underdense shell of width $s_-(t) \sim (t t_M)^{1/2}$ is created immediately following the overdense shell.  (Here we assume that the shell width grows proportionally to the scale factor, which is supported by the simulations.)  The density distribution around a newly formed black hole at $t\gg t_M$ is shown in Fig. \ref{fig:The-radiation-energy}(c).

The density contrast $\delta\equiv (\rho-\rho_r)/\rho_r$ is initially very large in the overdense shell and is ${\cal O}(1)$ in the underdense shell.  But it gets smaller as the shells expand, and at sufficiently late times we get into the regime where $|\delta|\ll 1$ and the shells can be treated as small perturbations, or sound waves.  We shall now discuss how the shells evolve, both in the regimes of strong shock and of sound waves.

\subsection{Strong shock}

The energy excess in the overdense shell can be estimated as\footnote{The energy in a spherically symmetric spacetime can be defined in terms of the so-called Misner-Sharp mass; see Section IV.D.}
\beq
E_+(t)  \sim  4\pi \delta\rho_+(t) R^2(t) s_+(t),                        
\eeq
where  $\delta\rho_+\equiv \rho_+ -\rho_r$, $\rho_+(t)$ is the average density of the shell, $R(t)$ is its radius, and
\beq
s_+(t) \sim s_{+i}\left(\frac{t}{t_i}\right)^{1/2}                                                                      
\eeq
is its thickness.  The shock is expected to dissipate as it propagates into the uniform radiation fluid. However, surprisingly, some early numerical work \cite{Liang, Anile} showed that for very strong shocks, the stronger the shock is the slower it damps. In particular, it was found that the damping rate of a planar shock in Minkowski spacetime initially grows with the increase of the shock strength $\delta$, but reaches a maximum at $\delta\sim 100$ and approaches zero in the limit of $\delta\to\infty$ \cite{Liang}.  For a strong planar shock in a radiation dominated universe, the decrease of $\rho_+(t)$ is mainly caused by the cosmological expansion.  It was shown in Ref. \cite{Anile}, that
\beq
\frac{\dot{\delta \rho_+}}{\delta \rho_+} \leq -\frac{2}{t}. 
\label{shock}
\eeq
The inequality is saturated in the limit of $\delta\to\infty$. In this case, $\delta\rho_+(t)$ redshifts in the same way as $\rho_r(t)$ does, which makes $\delta\approx {\rm const}$.\footnote{Note that in the simulation illustrated in Fig. \ref{fig:The-radiation-energy}(a) we have $\delta\sim 100$, which corresponds to the strongest damping of the shock.  The characteristic damping time in this regime is comparable to the thickness of the overdensed shell $s_+$. This explains why the density contrast $\delta$ is rapidly decreasing.  We were not able to explore much larger values of $\delta$, since that would require excessive amounts of computer time.}

For a large bubble, the initial shock radius is $R_i\gg t_i$.  It grows with the expansion of the universe as $R_s(t) \approx R_i (t/t_i)^{1/2}$ until it comes within the horizon at $t_H\sim R_i^2 /t_i$.  In this regime, the spherical shell can locally be approximated by a planar sheet and the results of Refs.~\cite{Liang,Anile} should apply.  From Eq.~(\ref{shock}) we have
\beq
\delta\rho_+(t)\lesssim \delta\rho_{+i}\left(\frac{t_i}{t}\right)^2 
\label{ineq1}
\eeq
and
\beq
E_+(t) \lesssim E_i \left(\frac{t_i}{t}\right)^{1/2}.
\label{ineq2}
\eeq
Then at the time of horizon crossing the energy excess in the shell satisfies
\beq
E_+ (t_H) \lesssim R_i^2/t_i \sim  t_H.
\label{ineq3}
\eeq
The inequalities (\ref{ineq1})-(\ref{ineq3}) are saturated in the limit of strong shock, $\delta\to\infty$.

Eq.~(\ref{ineq3}) tells us that the energy excess $E_+(t_H)$ does not exceed the energy contained in a horizon region of unperturbed FRW, $E_H \sim \rho_r t_H^3 \sim t_H$.  Black holes of mass $M>M_*$ have $M\sim t_H$; hence in this case $E_+(t_H)\lesssim M$, even though the initial energy of the shock is $E_i\gg M$.  On the other hand, for black holes with $M\ll M_*$ we may have $E_+(t_H) \gg M$.

The density contrast in the underdense shell at the time of horizon crossing is $|\delta_-(t_H)| \lesssim 1$, while $\delta_+(t_H)$ may be large or small, depending on the amount of shock damping. At $t>t_H$ the density contrast decreases and eventually becomes a small perturbation to the FRW background, where it can be treated as a sound wave.  We will show in the next subsection that the energy excess $E_+$ stays constant in this regime.

\subsection{Sound waves}

At large distances from the black hole the metric can be approximated as FRW metric,
\beq
ds^2 = dt^2 - a^2(t)\left(dr^2 + r^2 d\Omega^2\right),
\eeq
with $a(t)=(t/t_i)^{1/2}$.  The dynamics of a small spherically-symmetric density fluctuation $\delta(t,r)$ in this background is described by the equation~\cite{Silk:1967kq}
\beq
{\ddot\delta}+\frac{1}{2t}{\dot\delta}-\frac{1}{2t^2}\delta -\frac{c_s^2}{a^2}\left(\delta'' +\frac{2}{r} \delta'\right)=0.
\label{soundeq}
\eeq
where we take into account the self-gravitational effect. Here, overdots and primes stand for derivatives with respect to $t$ and $r$, respectively, and $c_s=3^{-1/2}$ is the speed of sound in a radiation-dominated plasma. We neglect the diffusion term that comes from the Silk damping effect, which we discuss qualitatively in Sec.~\ref{sec:mu-distortion}. 

The solution of Eq.~(\ref{soundeq}) for an outward propagating wave is
\beq
\delta(t,r)=At^{1/4} H^{(2)}_{3/2}\left(2k\sqrt{\frac{t_i t}{3}}\right)\frac{1}{r}e^{ikr} ,
\label{solution}
\eeq
where $A$ and $k$ are constants and 
\beq
H^{(2)}_{3/2}(z)=\sqrt{\frac{2}{\pi z}} e^{-iz} \left(\frac{i}{z} -1\right)
\eeq
is a Hankel function.  The argument of the Hankel function is the ratio of the sound horizon radius, $2c_s t$, to the wavelength of the perturbation, $\lambda(t) = a(t)/k$.  

At late times, when $\lambda \ll t$, we have 
\beq
\delta(t,r) \propto \frac{1}{r} \exp\left[ikr -2ik(t_i t/3)^{1/2}\right]. 
\label{latetime}
\eeq
In the overdense shell, waves of this form are superposed into a narrow wave packet of initial radius $R_i$ and width $k^{-1} \ll R_i$.  The width of the packet at time $t$ is $\sim a(t)/ k = (t/t_i)^{1/2} k^{-1}$ and its peak is at
\beq
r_s(t)=R_i +2\sqrt{\frac{t_i t}{3}}.
\eeq
The factor $1/r$ in Eq.~(\ref{latetime}) can then be approximately replaced in the packet by $1/r_s(t)$. Hence, at $t\ll R_i^2/t_i \sim t_H$, when $r(t)\approx {\rm const}$, the wave packet has a nearly constant amplitude\footnote{Note that this is the same behavior as we found for strong shocks with $\delta\gg 1$.  The reason is that the shock dissipation vanishes both in the limit of $\delta\to\infty$ and $\delta\to 0$.}, while at $t>t_H$ its amplitude decreases as $t^{-1/2}$. The energy carried by the expanding shell is 
\beq
E_+(t) \sim 4\pi k^{-1} r_s^2(t) a^3(t) \rho_r \delta_+(t,r_s) \propto t^{-1/2} r_s(t),
\label{Et}
\eeq
where $\rho_r(t)=3/32\pi t^2$ is the unperturbed FRW radiation density. This decreases as $t^{-1/2}$ at $t<t_H$ and remains constant at $t>t_H$. The above analysis should also be applicable to the underdense shell at times $t\gg t_H$.\footnote{Strictly speaking, the results of this subsection cannot be applied at $t>t_{\rm eq}$, since we used the radiation era expansion law $a(t)\propto t^{1/2}$ and the speed of sound $c_s=3^{-1/2}$ in a radiation-dominated plasma.  Analysis of sound waves at $t_{\rm eq} < t < t_{\rm rec}$ in Ref.~\cite{peebles1980large} shows that during this period the amplitude of the waves $\delta$ decreases compared to Eq.~(\ref{latetime}), but only by a factor ${\cal O}(1)$.  We will therefore use Eq.~(\ref{latetime}) to estimate $\delta$ at $t\sim t_{\rm rec}$ in Sec. V.}

\subsection{Energy considerations}

We cannot give an accurate analytic description of the evolution of overdense and underdense shells, except in the limiting cases of strong shock and of sound waves.  We can, however, make reasonable guesses about the energy carried by these shells after horizon crossing, $t>t_H$.  

\begin{figure}
   \centering
\includegraphics[scale=0.16]{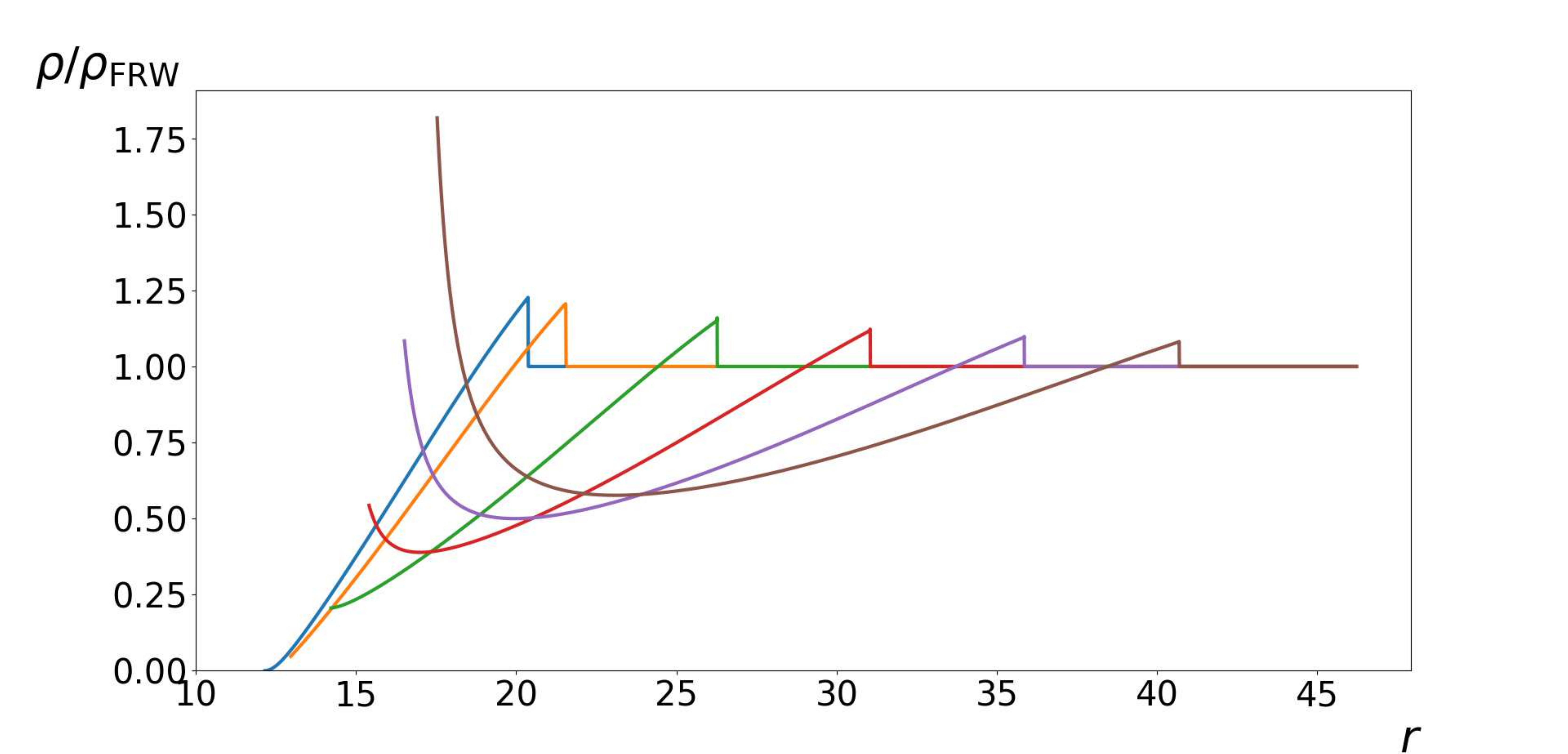}
\caption{\label{rho2}The radiation energy density profile at different moments in a supercritical case with $H_b=0.75H_{i}$, $\sigma \approx 0.003 H_{i}M_p^2,$ and $R_{i}=10H_{i}^{-1}$ [A similar plot in subcritical case is Fig. \ref{fig:The-radiation-energy}(b)]. The first (blue) and the second (orange) profiles respectively correspond to moments before and after the black hole formation, when $t\sim M \sim t_H$. The value of $\rho$ on the apparent horizon is negligible compared to the FRW density when the black hole is formed. Later it grows to a large value and the underdense shell propagates outwards following the shock.}
\end{figure}

For the analysis of $\mu$-distortion it will be sufficient to consider black holes with $M>M_*$ (see Sec. V).  From now on we shall therefore focus on this case.  A black hole with $M>M_*$ forms at $t\sim t_H \sim M$, and the underdense shell forms at about the same time.  Simulations suggest that the initial perturbation amplitude in such a shell is $\delta_- (t\sim t_H)\sim -1$ (see Fig. \ref{rho2}), and the energy deficit in the shell is 
\beq
E_- \sim \rho_r \delta_- t_H^3 \sim -M.
\label{deficit}
\eeq
(Note that at $t\sim t_H$ the thickness of the underdense shell is comparable to its radius $t_H$.) The total energy deficit in the two shells must compensate for the black hole mass,
\beq
E_+ +E_- \sim -M.
\label{E+E-}
\eeq
Hence we conclude that the energy excess in the overdense shell must satisfy
\beq
E_+ \lesssim M,
\label{E+}
\eeq
in agreement with Eq.~(\ref{ineq3}).

At $t\gg t_H$, Eqs.~(\ref{E+E-}) and (\ref{E+}) should still apply, and thus we must have
\beq
E_-(t) \sim \rho_r(t) \delta_-(t) R_s^2(t) s_-(t) \sim -M.
\label{deficit}
\eeq
and
\beq
\delta_-(t) \sim (M/t)^{1/2} \ll 1.
\eeq
Here, $R_s(t)\approx 2c_s t$ is the shell radius (the sound horizon) and $s_-(t)\sim (M t)^{1/2}$ is its thickness. Since $\delta_- \ll 1$, the results of Sec.~IV.B should apply and the total energy of the shell should remain constant, in agreement with Eq.~(\ref{deficit}).

\subsection{Numerical simulations}

In Ref.~\cite{Deng:2017uwc} Einstein's equations were solved numerically to study the evolution of radiation and spacetime outside the bubble. Here we use the same numerical code to verify the scenario of shock evolution outlined in subsections A-C.

The line element in a spherically symmetric spacetime can always be cast in the form
\begin{equation}
ds^{2}=A^{2}dt^{2}-B^{2}dr^{2}-R^{2}d\Omega^{2},
\end{equation}
where $A,B$ and $R$ are functions of the coordinates $t$ and $r$.  Radiation is regarded as a perfect fluid with energy density $\rho$ and pressure $P=\rho/3$. We chose a gauge where the spatial components of the fluid's four-velocity vanish, $u^\mu = (A^{-1},0,0,0)$. We assume that the fluid is completely reflected by the bubble, which implies that the bubble wall is comoving with the fluid.  This is rather convenient for the numerical implementation, because the wall can now serve as the inner boundary at a fixed $r$.  We used Israel's junction conditions to match the de Sitter spacetime inside the bubble and the spacetime outside, which gives the boundary condition on the wall.

\begin{figure}
   \centering
\includegraphics[scale=0.23]{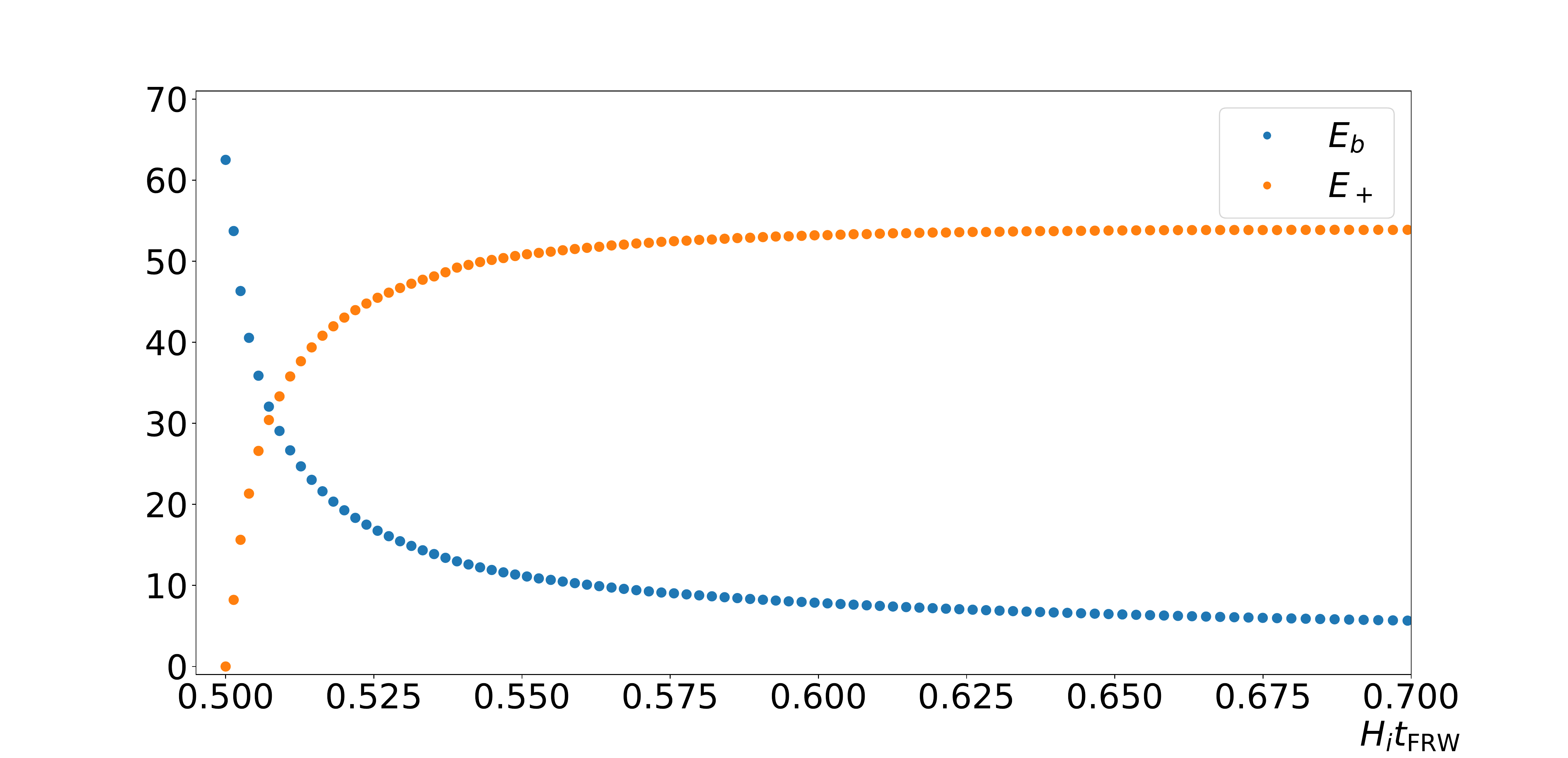}
\caption{\label{bubblemass}
The energy of the bubble $E_b$ and of the overdense shell $E_+$ in units of $M_p^2/H_i$ at early times for the same parameter values as in Fig.~\ref{fig:The-radiation-energy}. The bubble expands with an initial energy $E_i=H_i^2 R_i^3 / 2$ ($=62.5M_p^2/H_i$ in this example). Most of this energy is transferred to the shell within about a Hubble time.}
\end{figure}

Since it is assumed that thermalization takes place instantaneously after inflation ends, the resulting radiation is initially set to be homogeneous in the simulation, while the bubble expands with a large Lorentz factor. It can be seen in Fig. \ref{fig:The-radiation-energy}(a) that, as the bubble wall hits the ambient radiation, an overdense shell develops and turns into a shock front within a very short time. During this time, the bubble loses most of its energy, which is transferred to the shock (Fig. \ref{bubblemass}). Restricted by the capability of our code, we didn't observe the effect of strong shock discussed in Subsection A. In all our simulations the shock dissipates quickly and eventually approaches $\delta \propto t^{-1/2}$ as expected in Subsection B. 

As the shock propagates away, a shell with low density surrounding the bubble is left behind. In the supercritical case, a wormhole then develops in this underdense region as the bubble grows exponentially into the baby universe. To avoid code crash, we then remove the wall as well as a small layer outside the wall, which should not affect the evolution of the parent universe.

The wormhole later turns into two black holes, one for observers in the parent universe and the other in the baby universe. The black hole formation is signaled by the appearance of an apparent horizon, and the black hole mass can be determined from the apparent horizon radius. The two black holes are identical at the beginning, but subsequent accretion would give them different masses. To avoid simulation crash, the black hole region is cut off at the apparent horizon. Moreover, since we are mainly interested in how the FRW region is perturbed, the baby universe is also discarded. After black hole formation, the underdense region begins to fill up. In particular, as the wormhole "pinches off", a singularity arises, and radiation density near the black hole begins to grow from almost zero to a large value.

For the purposes of our numerical simulations, it is useful to have a precise definition of what we mean by energy in our dynamical spacetime.  The energy contained in a sphere of area-radius $R$ can be identified with the Misner-Sharp mass \cite{Misner:1964je,Hayward:1994bu}, which in our coordinates is given by
\begin{equation}
M_{\rm MS}(R) = \frac{R}{2}\left(1-\frac{R^{\prime 2}}{B^2}+\frac{\dot{R}^{2}}{A^2}\right),
\end{equation}
where prime and dot stand respectively for derivatives with respect to $r$ and $t$, as before.  Well after the black hole is formed, at $t\gg t_H$, the radius of its apparent horizon is $R=2M$ and the Misner-Sharp mass within that radius is equal to the black hole mass, $M_{\rm MS}(2M)=M$.  Another example is the cosmological horizon, $R=2t$.  At $t\gg t_H$ it is in the unperturbed FRW region, where $A=1$, $B=(t/t_i)^{1/2}$, $R=r(t/t_i)^{1/2}$, and the Misner-Sharp mass is $M_{\rm MS}(2t)= t$.  Note that the same value of the mass can be obtained from 
\beq
 4\pi  \rho_r(t) \int_0^{2t} R^2 dR = t
\label{pure}
\eeq
in a pure FRW universe. 

It can be shown that
\begin{equation}
M_{\rm MS}^{\prime}(R) = 4 \pi \rho R^2 R^{\prime}.
\end{equation}
Integrating this equation from $R=2M$ to $R=2t$, we obtain
\beq
4\pi \int_{2M}^{2t} \delta\rho R^2 dR = -M \left(1-\frac{M^2}{t^2}\right),
\label{integral}
\eeq
where $\delta\rho=\rho-\rho_r$, as before, and we have used Eq.~(\ref{pure}).  Since $\delta\rho=0$ outside the shock front, we can extend the integration in (\ref{integral}) to $R=\infty$. Also, at $t\gg M$, the term $M^2/t^2$ in the parentheses can be neglected, and we obtain
\beq
4\pi \int_{2M}^\infty \delta\rho R^2 dR \approx -M.
\label{energyconservation}
\eeq
Assuming that $\delta\rho$ is significantly different from zero only within the two expanding shells, this can be interpreted in the sense that the black hole mass is compensated by the energy deficit in the shells, as in Eq.~(\ref{E+E-}).  The energy of the overdense shell can be defined as $E_+ = M_{\rm MS}(R_f)-M_{\rm MS}(R_0)$, where $R_f$ is the radius of the shock front and $R_0$ is the radius where $\delta\rho$ changes sign.  The energy of the underdense shell $E_-$ can be similarly defined.

\begin{figure}
   \centering
\includegraphics[scale=0.23]{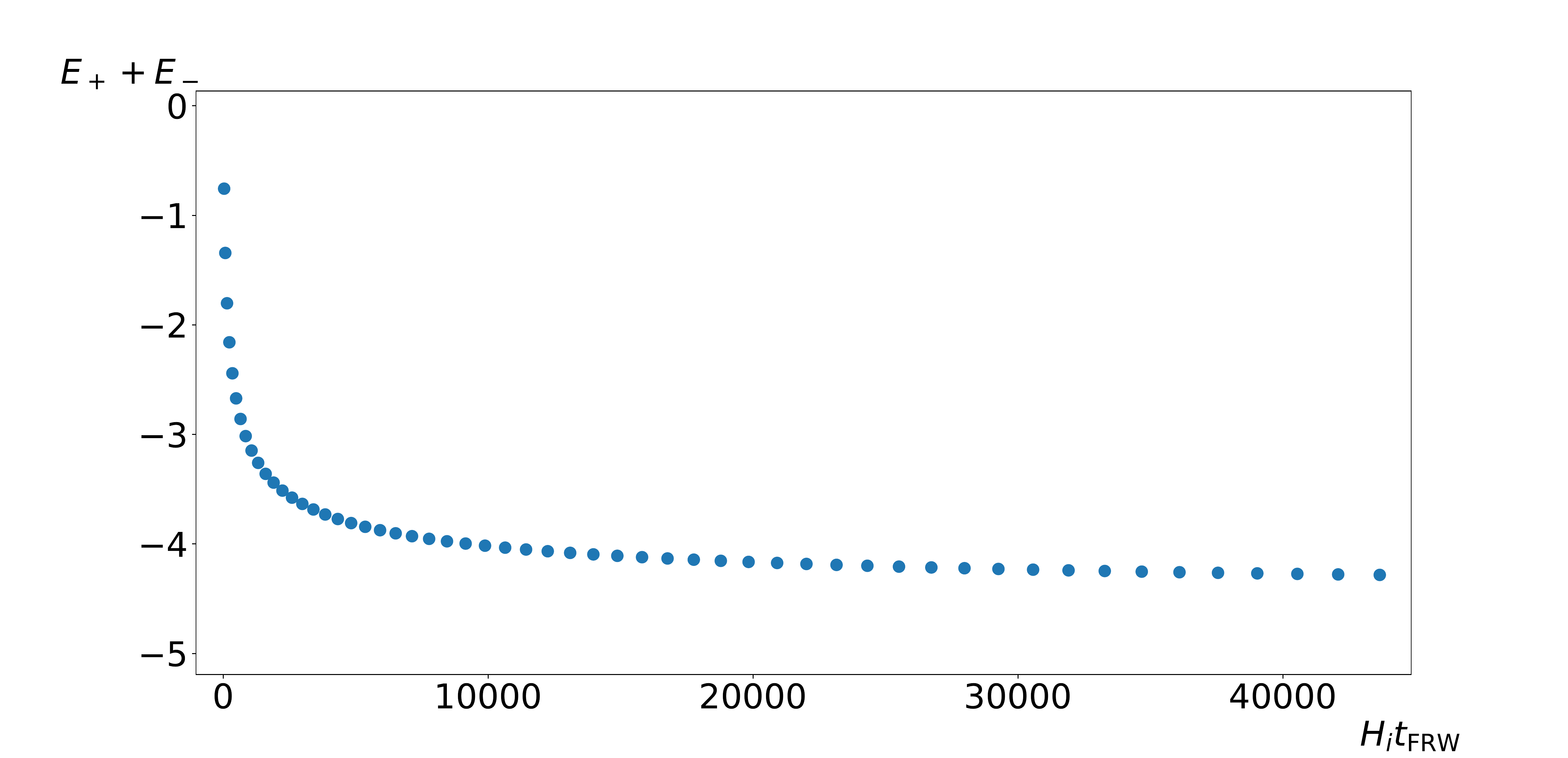}
\caption{\label{E+E-plot}
The combined energy of the two shells $E_++E_-$ in units of $M_p^2/H_i$ for the same parameter values as in Fig.~\ref{fig:The-radiation-energy}. At late times this energy approaches a constant $\sim -M$ ($M\approx 4M_p^2/H_i$ in this example).}
\end{figure}

Eq.~(\ref{E+E-}) can be expected to hold only approximately because the black hole mass grows with time due to accretion and the radiation density differs from FRW near the black hole.  However, both of these effects become smaller with time, so we expect Eq.~(\ref{E+E-}) to become increasingly accurate. We have verified that the combined energy in the two expanding shells approaches a constant $\sim -M$ at late times (see Fig. \ref{E+E-plot}).  At the same time, the density in the region between the shells and the black hole approaches that in the unperturbed exterior FRW region (Fig. \ref{fig:The-radiation-energy}(c)), except in the immediate vicinity of the black hole.  All our simulation results are consistent with the shock evolution scenario we outlined in subsections A-C.  In the next section we shall discuss how the energy deficit (and excess) in the expanding shells may lead to a spectral distortion in the CMB.

\section{Photon diffusion and $\mu$-distortion}
\label{sec:mu-distortion}

\subsection{Photon diffusion}

Sound waves in a radiation-dominated plasma are dissipated by photon diffusion with a characteristic length scale
\beq
\lambda_D(t) \sim (t/\sigma_T n_e)^{1/2} \propto t^{5/4}.
\eeq 
Here, $\sigma_T$ is the Thompson cross-section and $n_e$ is the electron density.  This dissipation process was first studied by Silk~\cite{Silk:1967kq} and is known as Silk damping. The diffusion length is initially small compared to shell thickness $s_\pm(t)$, but it grows faster and at some point catches up with $s_+$ and then with $s_-$.  The shells are then smeared by photon diffusion and eventually turn into a single shell of thickness $s_M(t)\sim\lambda_D(t)$ and energy deficit $E_M\sim -M$, where the subscript $M$ indicates that the shell is centered at a black hole of mass $M$.  As $\lambda_D$ grows, the density contrast in the shell decreases proportionately, so that the total energy deficit remains the same as before.  So, in this regime
\beq
s_M(t) \sim \lambda_D(t), 
\label{sM}
\eeq
\beq
\delta_M (t)\sim \frac{M}{\rho_r \cdot 4\pi (2c_s t)^2 \lambda_D} \sim \frac{2M}{\lambda_D}.
\label{deltas}
\eeq

An underdense shell of thickness $\sim \lambda_D$ includes a superposition of sound waves of wavelength $\lambda\gtrsim \lambda_D$, with most of the energy being in waves of $\lambda\sim\lambda_D$.  These waves are dissipated into heat in about a Hubble time.  The remaining, less energetic waves of longer wavelength make up a thicker shell with a smaller density contrast, as can be seen from Eq.~(\ref{deltas}).  This picture is supported by an explicit solution of equations of motion for a viscous fluid in Appendix~\ref{sec:sound wave energy}.

We also show in the Appendix that the energy density of sound waves which is dissipated by Silk damping is $\rho_s = \frac{1}{4}\rho_r \delta^2$. Hence the sound wave energy in a spherical shell around a black hole of mass $M$ can be estimated as
\beq
E_{Ms}(t) \sim \frac{1}{4} \rho_r(t) \delta_M^2(t) V_s(t),
\label{EM}
\eeq
where
\beq
V_s(t) \sim 4\pi (2c_s t)^2 \lambda_D(t)
\label{Vs}
\eeq
is the shell volume, $2c_s t$ is its radius, and $\lambda_D(t)$ is its thickness. Combining Eqs.~(\ref{EM}), (\ref{Vs}) and (\ref{deltas}), we obtain
\beq
E_{Ms}(t) \sim \frac{M^2}{2 \lambda_D(t)}.
\label{EMs2}
\eeq

\subsection{$\mu$-distortion}
\label{mu-distortion}

The radiation temperature inside the underdense shell is lower than the temperature outside; hence photon diffusion produces a mixture of photons at different temperatures and distorts the Planck spectrum.  At $t< t_{\rm th} \sim 7\times 10^6 s$ this mixture is completely thermalized and can be fully characterized by its temperature, which differs from the background temperature $T$ by the amount $\delta T/ T \sim \frac{1}{4} \delta_M(t)$.   At $t_{\rm th}<t<t_\mu \sim 3\times 10^9 s$, the photon number changing processes (like bremsstrahlung and double Compton scattering) are ineffective, but partial thermalization still occurs, resulting in a nonzero chemical potential for the photons \cite{Sunyaev:1970er,Daly, Barrow_1991}.

The spectral distortion produced by an expanding underdense shell at time $\sim t$ is confined to a spherical region of sound horizon radius $(\sim 2c_s t)$, with a sphere of comoving radius $\sim 2c_s t_{\rm th}$ cut out.  As the shell propagates outwards, the distortion extends to larger comoving radii.  In the meantime, the $\mu$-distorted spectrum spreads beyond the initial regions where it was produced by photon diffusion.  The comoving scales of the sound horizon during the $\mu$-era range from $\sim 0.1$~Mpc to $\sim 2$~Mpc, while the Silk damping scale continues to grow until the time of recombination, reaching the comoving value $\lambda_D^{(c)}(t_{\rm rec}) \sim 10$~Mpc (with the superscript $(c)$ indicating that this is a comoving scale).  We shall refer to $\lambda_D^{(c)}(t_{\rm rec})$ as the Silk scale and to a comoving region of this size as a Silk region. The Silk scale is significantly larger than the comoving shell radius during the $\mu$-era, which means that the damping of sound waves acts effectively as a pointlike energy release.  Hence photons with a distorted spectrum induced by an underdense shell propagating outwards from a black hole are eventually mixed with the background photons within a Silk region centered at the black hole.

In the context of $\mu$-distortion, the chemical potential is traditionally defined as $\mu\equiv -\mu_{\rm th}/T$, where $\mu_{\rm th}$ is the thermodynamic chemical potential. If the black hole density at $t_{\rm rec}$ is $n_M(t_{\rm rec}) \gg \lambda_D^{-3}(t_{\rm rec})$, then photons affected by different shells mix together, resulting in a uniform chemical potential $\mu$.  In this case $\mu$ can be estimated as~\cite{Sunyaev:1970er, Daly, Barrow_1991, Hu:1992dc, Chluba:2012we}
\beq
\mu_M \sim -\int_{t_{\rm th}}^{t_\mu} dt \frac{d}{dt} \left(\frac{\rho_{Ms}}{\rho} \right) = \frac{\rho_{Ms}}{\rho_r} (t_{\rm th}) - \frac{\rho_{Ms}}{\rho_r} (t_\mu).
\label{mudistortion}
\eeq
Here, $\rho_{Ms}$ is the sound wave energy density in shells centered at black holes of mass $\sim M$ averaged over a large volume,
\beq
\rho_{Ms}(t) \sim E_{Ms}(t) n_M(t) ,
\eeq
where $E_{Ms}(t)$ is from Eq.~(\ref{EMs2}) and
\beq
n_M(t)\sim \rho_{\rm CDM}(t)\frac{f(M)}{M} \sim \Gamma (Mt)^{-3/2}
\eeq
is the number density of black holes of mass $\sim M$ during the radiation era (assuming that $M>M_*$).  Thus we have
\beq
\frac{\rho_{Ms}}{\rho_r} (t) \sim 10 \Gamma \frac{ (M t)^{1/2}}{ \lambda_D (t)} \propto t^{-3/4}. 
\label{mudistortion1}
\eeq
Contributions to $\mu$ due to black holes in different mass ranges simply add up. 

In the above analysis we assumed that the thickness of the expanding underdense shell gets comparable to the photon diffusion length prior to $t = t_{\rm th} \sim 7\times 10^6 s$.  This holds only for PBH with $s_- (t_{th})\lesssim \lambda_D(t_{\rm th})$, or $M\lesssim \lambda_D^2(t_{\rm th})/t_{\rm th} \sim 10^5 M_\odot$.   For larger black holes, the overdense shell with energy excess $E_+ \sim M$ is smeared to a thickness $s_+\sim \lambda_D$.  Then the density contrast $\delta_+$ in this shell is given by Eq.~(\ref{deltas}).  The underdense shell is much thicker, $s_-\gg s_+$, and has a much smaller density contrast, $\delta_- \sim (s_+/s_-) \delta_+$, so the sound wave dissipation occurs mainly in the overdense shell.  The sound wave energy can still be estimated by Eq.~(\ref{EMs2}), and thus \eq{mudistortion1} can still be used to give $\mu$. 

It follows from Eq.~(\ref{mudistortion1}) that the largest $\mu$-distortion (at a fixed $M$) is produced during the earliest part of the $\mu$-era, $t\sim t_{\rm th}$:
\beq
\mu_M \sim 10 \Gamma \frac{ (M t_{\rm th})^{1/2}}{ \lambda_D (t_{\rm th})}.
\label{mutth}  
\eeq
We note also that the magnitude of $\mu$ in Eq.~(\ref{mutth}) grows with the black hole mass.  This means that the dominant contribution to the $\mu$-distortion in a given Silk region is due to the largest black hole contained in that region.  The corresponding black hole mass ${\bar M}$ can be found from Eq.~(\ref{nM}) by setting $n_M \sim (\lambda_D^{(c)} (t_{\rm rec}))^{-3} \sim (10 \, {\rm Mpc})^{-3}$, which gives ${\bar M} \sim 10^{15} \Gamma^{2/3} M_\odot$.  Substituting $M\sim {\bar M}$ in (\ref{mutth}) we find 
\beq
\mu_{\bar{M}} 
\sim 10^6 \Gamma^{4/3}. 
\label{barmu}
\eeq 

A typical Silk region will contain one or few black holes of mass $\sim {\bar M}$.  Since the number of such black holes will vary significantly from one region to another, there will be large fluctuations of $\mu$ on the comoving Silk scale $\lambda_D (t_{\rm rec})$.  Thus we expect
\beq
{\bar\mu} \sim \delta\mu_{\rm rms} \sim \mu_{\bar M},
\eeq
where ${\bar\mu}$ and $\delta\mu_{\rm rms}$ are respectively the mean value and the rms fluctuation of $\mu$. With $\Gamma\lesssim 10^{-12}$, both the average value ${\bar\mu}$ and the fluctuations are smaller than $\mu\sim 10^{-8}$, which is expected from Silk damping of sound waves in the standard $\Lambda$CDM model~\cite{Chluba:2016bvg}.

There will, however, be some rare regions containing black holes with $M\gg {\bar M}$.  Their probability distribution is
\beq
dP(M>{\bar M})\sim \left(\frac{\bar M}{M}\right)^{3/2} \frac{dM}{M}. 
\label{dP}
\eeq
The $\mu$-distortion in such regions will be anomalously large.  It can be estimated by comparing Silk regions with the largest black hole masses $M\sim {\bar M}$ and $M\gg{\bar M}$.  In both cases the $\mu$-distortion is due to a single shell and is proportional to the sound wave energy (\ref{EMs2}) in that shell at $t\sim t_{\rm th}$. This energy is proportional to $M^2$; hence we must have 
\beq
\mu_M \sim \bar{\mu} \lmk \frac{M}{\bar{M}} \rmk^2. 
\label{muM}
\eeq

Black holes that can induce a $\mu$-distortion in the CMB are located within a Silk distance of the LSS. The mass of the largest black hole we can expect to find in this range can be estimated from Eq.~(\ref{nM}) by setting $n_M \sim [4 \pi (10\, {\rm Gpc})^2 \times (10\,{\rm Mpc})]^{-1}$, which gives $M_{\rm max}\sim 10^5{\bar M}$.  The $\mu$-distortion in a Silk patch containing such a black hole is 
\beq
\mu_{\rm max}\sim 10^{10}{\bar\mu}.
\label{mumax}
\eeq

At $t>t_\mu$, scattering on electrons results in a Compton-distorted spectrum with a $y$-parameter given by
\beq
y_M \sim -\int_{t_\mu}^{t_{\rm rec}} dt \frac{d}{dt} \left(\frac{\rho_{Ms}}{\rho} \right) \approx \frac{\rho_{Ms}}{\rho_r} (t_\mu) 
\label{ydistortion}
\eeq
Once again, the main contribution comes from the earliest part of the $y$-era, $t\sim t_\mu$.  The resulting distortion is smaller than the $\mu$-distortion by the factor $(t_{\rm th}/t_\mu)^{3/4} \sim 0.01$. Since this is subdominant, we do not discuss the $y$-distortion further in this paper.

In most of this paper we considered black holes with $M>M_*$.  For $M<M_*$ we do not have a reliable estimate of $E_\pm$, and the shells may in principle have energies $|E_\pm| \gg M$.  We note, however, that the values of $E_\pm$ are not important for $M\lesssim 10^5 M_\odot$, in which case the two shells merge with a total energy deficit $E_+ + E_- \sim -M$, so  Eq.~(\ref{deltas}) can be used to estimate the density contrast.  Hence our results should apply as long as $M_*\lesssim 10^5 M_\odot$, which is the case we are mostly interested in (see Eq.~(\ref{Planck satellite constraint})).

\subsection{Observational constraints}

The current upper limits on the mean value and the rms fluctuation of the $\mu$-distortion from COBE-FIRAS and Planck data are ${\bar\mu} < 9\times 10^{-5}$ and $\delta\mu_{\rm rms} < 6\times 10^{-6}$, respectively (see \cite{Khatri:2015tla} and references therein).  The values predicted by our model, ${\bar\mu}\sim\delta\mu_{\rm rms} \sim 10^6 \Gamma^{4/3}$ are well below these observational bounds for bubble nucleation rates $\Gamma \lesssim 10^{-12}$.  A much more stringent constraint comes from our prediction of a few Silk patches with a much larger spectral distortion, $\mu_{\rm max}\sim 10^{10}{\bar\mu}$.  The angular size of these patches on the microwave sky is $\sim 10'$.

Figure~3 in Ref.~\cite{Khatri:2015tla} shows the probability distribution $P(\mu)$ of the largest $\mu$-distortion allowed by the Planck data.  More precisely, $P(\mu)$ is the fraction of pixels (of size $\sim 10'$) where $\mu$ can exceed a given value.  The value of $\mu\sim \mu_{\rm max}$ is expected in only one or few pixels, which corresponds to $P(\mu)\sim 10^{-6} - 10^{-7}$, and it follows from the figure that the value of $\mu$ in such pixels is bounded by $\mu\lesssim 10^{-4}$.  Eq.~(\ref{mumax}) then yields a bound on the nucleation rate,
\beq
\Gamma\lesssim 10^{-15}.
\label{strongbound}
\eeq
We note that the few pixels containing the largest black holes with $M \sim M_{\rm max}$ may be obscured on the sky (e.g., by our galaxy).  A black hole with $M \sim M_{\rm max}/3$ would give $\mu(M) \sim 0.1 \mu_{\rm max}$.  The resulting bound on $\Gamma$ would then be weaker by a factor of $10$.\footnote{The angular size of these patches is $\sim 10' \sim 0.2^\circ$ while PIXIE will have an angular resolution of $\Delta \theta \simeq 1.6^\circ$~\cite{Abitbol:2017vwa}. Hence the largest distortion that can be measured by this experiment is $(0.2/1.6)^2 \mu_{\rm max}$. Its sensitivity would reach $\abs{\mu} \simeq 3.6 \times 10^{-7}$ for the spatially-constant component of $\mu$, so we expect that the sensitivity of one pixel is about $\sqrt{2l (1.6^\circ) +1} \sim 10$ times weaker than this proposed sensitivity, where $l (1.6^\circ)$ represents the angular moment corresponding to $1.6^\circ$. This implies that we would observe $\mu_{\rm max}$ in the near future if $\mu_{\rm max} \gtrsim 10^{-4}$. However, this is as large as the present constraint by the Planck data. 
}

The bound (\ref{strongbound}) excludes the value of $\Gamma\sim 10^{-12}$, which is needed to account for LIGO observations, but is consistent with the condition $\Gamma\gtrsim 10^{-17}$, which is necessary for seeding SMBHs. The LIGO mergers may still be accounted for if we relax the assumption that the bubble nucleation rate $\Gamma$ remains constant during inflation.  As we mentioned in Section II, this assumption is justified for small-field models of inflation, where the inflaton field displacement during the slow roll is rather small, $\Delta\phi\ll M_p$.  On the other hand, in large-field models the inflaton traverses a large distance in the field space, $\Delta\phi \gtrsim M_p$, and the nucleation rate may change by many orders of magnitude.  In this case, the nucleation rate $\Gamma$ in the mass distribution function (\ref{fM1}), (\ref{fM2}) is a function of $M$.  It is possible, in particular, that $\Gamma$ remains nearly constant for the range of masses relevant for LIGO and SMBH seeds, but declines significantly at $M\sim M_{\rm max}$.  The bound (\ref{strongbound}) can be avoided if the mass distribution is effectively cut off at
\beq
M_c \lesssim 3\times 10^{-6} \Gamma^{-2/3} {\bar M}.
\label{Mc}
\eeq
We could have, for example, a nearly constant $\Gamma(M) \sim 10^{-12}$ with a cutoff at some mass $M_c$ anywhere in the range $10^7 M_\odot \lesssim M_c \lesssim 10^{10} M_\odot$.  Then the PBHs formed by our mechanism can account for both LIGO and SMBH observations, with the SMBH seeds having masses $M_{\rm seed}\sim 10^6 M_\odot$.  The largest black hole in a typical Silk patch is then ${\bar M}\sim 10^7 M_\odot$, and the typical $\mu$-distortion is ${\bar\mu} \sim 10^{-10}$.  The largest distortion will be reached in patches containing black holes of mass $\sim M_c$, $\mu_{M_c}\sim {\bar\mu} (M_c/{\bar M})^2 \lesssim 10^{-5}$.  

We finally consider the temperature fluctuations induced by the expanding shells in the CMB.  At $t\sim t_{\rm rec}$ the shells radii are set by the sound horizon, which corresponds to the angular scale of $\sim 1^\circ$, and have thickness $\sim 10'$.  The temperature fluctuation in such a shell is 
\beq
\frac{\delta T}{T} \sim \frac{1}{4} \delta_M(t_{\rm rec}) \sim \frac{M}{2\lambda_D (t_{\rm rec})}.
\label{dT/T}
\eeq
If this is large enough, the fluctuations caused by the largest black holes could induce ring-like temperature patterns in the CMB sky.  However, with the mass bounded by the cutoff (\ref{Mc}), we have $\delta T/T\lesssim 10^{-8}$. This is much smaller than the rms CMB fluctuations $(\sim 10^{-5})$; hence these temperature fluctuations are unobservable.

\section{Conclusions and discussion}

We have estimated the CMB spectral distortions expected in the scenario of primordial black hole formation by vacuum bubbles nucleated during inflation.  When inflation ends, the bubbles run into the ambient plasma, producing strong shocks followed by underdensity waves, which propagate outwards.  The bubble themselves eventually form black holes with a wide distribution of masses.  These black holes may serve as seeds for supermassive black holes observed at galactic centers if the bubble nucleation rate during inflation (per Hubble volume per Hubble time) satisfies $\Gamma\gtrsim 10^{-17}$ and may account for LIGO observations if $\Gamma\sim 10^{-12}$.   The latter value is marginally consistent with the Planck satellite constraint, $\Gamma\lesssim 10^{-12}$.

The expanding shocks and underdensities are eventually smeared by photon diffusion, resulting in a $\mu$-type distortion of the CMB spectrum.  We found that the magnitude of this distortion averaged over the sky is of the order
\beq
{\bar\mu} \sim 10^6 \Gamma^{4/3}.
\eeq
With $\Gamma\lesssim 10^{-12}$, this is too small to be observed.  We also found that the values of $\mu$ in this scenario are highly variable over the sky, $\delta\mu\sim \bar\mu$, with a typical angular scale of variation $\sim 10'$, corresponding to the Silk scale at the time of recombination.  The distortion in a given Silk-size region is mainly due to the largest black hole that the region contains; its typical mass is ${\bar M} \sim 10^{15} \Gamma^{2/3} M_\odot$.  Hence we expect ${\cal O}(1)$ fluctuations from one region to another.  Moreover, some rare Silk patches will contain black holes of mass $M\gg {\bar M}$, with a probability distribution given by Eq.~(\ref{dP}).  The values of $\mu$ will therefore have localized peaks with a probability distribution
\beq
dP(\mu) \propto \frac{d\mu}{\mu^{7/4}},
\eeq 
ranging from $\sim {\bar\mu}$ to ${\mu_{\rm max}} \sim 10^{10} {\bar\mu}$.  This spiky distribution of the spectral distortion is a unique observational signature of our black hole formation model.

The maximal distortion $\mu_{\rm max}$ is expected to be attained in only one or few Silk-size patches of the sky.  Planck observations impose strong limits on such isolated spikes \cite{Khatri:2015tla}, resulting in a bound on the bubble nucleation rate,  $\Gamma\lesssim 10^{-15}$.  This rules out the value of $\Gamma\sim 10^{-12}$, which is needed to account for LIGO observations.  The bound, however, can be avoided if we relax the assumption that the nucleation rate $\Gamma$ remains constant during inflation.  A variable $\Gamma$ can naturally arise in models of large-field inflation, where the inflaton field traverses a large distance $\Delta\phi\gtrsim M_p$ in the field space during the slow roll.

Another simplifying assumption that we made in this paper is that radiation is completely reflected from the bubble wall.  If reflection is incomplete, the shock wave and the resulting spectral distortion would be weaker and the bound on $\Gamma$ would be relaxed.  As a limiting case, one could consider a model where the bubble interacts with radiation only gravitationally.  We note finally that a wide distribution of PBHs can also be produced in a closely related scenario, where the black holes are formed by spherical domain walls nucleating during inflation \cite{Garriga:2015fdk, Deng:2016vzb}.  In this case, the walls do not have large Lorentz factors and do not produce strong shocks, but underdensity waves compensating for the black hole mass would still be formed.  We leave the analysis of spectral distortion in these scenarios to future work.

We have briefly discussed the possibility that the black hole distribution predicted in our scenario may lead to early formation of massive dark matter halos.  For bubble nucleation rates $\Gamma\gtrsim 10^{-15}$, the predicted density of halos with $M_{\rm halo}\gtrsim 10^{11} - 10^{12}M_\odot$ significantly exceeds that in the standard hierarchical structure formation model.  This excess of massive halos may account for some recent observations \cite{Kang:2015xhf, Steinhardt:2015lqa, Franck:2016aa}.

\begin{acknowledgments}
We are grateful to Yacine Ali-Haimoud, Ruth Daly, Andrei Gruzinov, and Jim Peebles for stimulating discussions and to Rishi Khatri, Avi Loeb and Misao Sasaki for useful comments.  This work was supported in part by the National Science Foundation under grant 1518742. HD is supported by the John F. Burlingame Graduate Fellowships in Physics at Tufts University.

\end{acknowledgments}

\appendix

\section{Photon diffusion in a plane sound wave pulse}
 
In this Appendix we study the effect of photon diffusion on a propagating sound wave pulse and estimate the energy dissipated in this process and the resulting $\mu$-distortion.

\subsection{Dissipated sound wave energy}
\label{sec:sound wave energy}

The thickness of the expanding shell we are interested in is small compared to its radius and to the horizon; hence it can be locally approximated by a plane sound wave pulse propagating in a radiation fluid $(P=\rho/3)$ in flat spacetime. The energy-momentum tensor of the wave is 
\beq
T^{00} = \rho_r (1+\delta) + \frac{4}{3}\rho u^2,
\label{T00}
\eeq
\beq
T^{i0} = \frac{4}{3}\rho_r (1+\delta)u^i,
\eeq
\beq
T^{ij}= \frac{1}{3}\delta_{ij}\rho_r (1+\delta) +\frac{4}{3}\rho_r (1+\delta)u^i u^j ,
\eeq
where $\rho_r={\rm const}$ is the unperturbed energy density, $u^i \ll 1$ is the fluid velocity, and $u^2=|u^i u_i|$. From energy-momentum conservation, $\partial_\nu T^{\mu\nu}=0$, we have in the linear approximation in the perturbation
\beq
{\dot\delta} +\frac{4}{3}\partial_i u^i = 0,
\label{conservation}
\eeq
\beq
\frac{4}{3}{\dot u}^i + \frac{1}{3}\partial_i \delta=0.
\label{conservation2}
\eeq
Combining these two equations we obtain the wave equation for the density perturbation,
\beq
{\ddot\delta}-c_s^2\nabla^2\delta=0, 
\label{waveeq}
\eeq
where $c_s^2 = 1/3$. 

For a plane wave pulse propagating in the $x$-direction, the solution of Eq.~(\ref{waveeq}) is an arbitrary function $\delta(x-c_s t)$. Then it follows from (\ref{conservation}) that
\beq
u^x = \frac{3}{4} c_s\delta,
\label{udelta}
\eeq
and Eq.~(\ref{T00}) becomes
\beq
T^{00} = \rho_r \left(1+\delta + \frac{3}{4} c_s^2 \delta^2\right).
\label{T001}
\eeq

In the presence of photon diffusion, the energy-momentum tensor acquires an extra term $\Delta T_{\mu\nu}$.  For an irrotational flow its divergence is given by \cite{Pajer:2013oca} 
\beq
\partial_\nu \Delta T^{0\nu}=0 ,
\eeq
\beq
\partial_\nu \Delta T^{i\nu} = - \frac{4}{3} \eta\nabla^2 u^i ,
\eeq
where the viscous coefficient $\eta$ is
\beq
\eta=\frac{16}{45}\rho_r \tau_\gamma
\eeq
and $\tau_\gamma = (\sigma_T n_e)^{-1}$ is the photon mean free path.  With this extra term, Eq.~(\ref{conservation2}) is modified,
\beq
\frac{4}{3}{\dot u}^i + \frac{1}{3}\partial_i \delta -\frac{4\eta}{3\rho_r} \nabla^2 u^i=0 ,
\label{conservation3}
\eeq
while Eq.~(\ref{conservation}) remains unchanged.  The wave equation for $\delta$ now takes the form
\beq
{\ddot\delta}-c_s^2\nabla^2\delta -\frac{\eta}{\rho_r} \nabla^2 {\dot\delta}=0 .
\label{waveeq2}
\eeq

Eq.~(\ref{waveeq2}) is easily solved in the Fourier representation.  In the limit of small viscosity, we find
\beq
\delta_k(t) = A_k \exp\left(-ic_s kt -\frac{k^2 \eta}{2\rho_r} t\right).
\eeq
To illustrate the effect of photon diffusion on the propagation of a flat pulse, let us consider a Gaussian pulse of initial width $s_0$, 
\beq
\delta(x,0)=\delta_0 \exp\left(-\frac{x^2}{2s_0^2}\right).
\eeq
Then
\beq
A_k = \frac{s_0\delta_0}{\sqrt{2\pi}} \exp\left(-\frac{k^2 s_0^2}{2}\right)
\eeq
and
\bea
\delta(x,t) &=& s_0\delta_0 \int\frac{dk}{\sqrt{2\pi}} e^{ik(x-c_s t)} e^{-\frac{k^2}{2}\left(s_0^2 +\frac{\eta}{\rho_r} t\right)}
\\
&=& \frac{\delta_0 s_0}{s(t)}\exp\left[-\frac{(x-c_s t)^2}{2s^2(t)}\right], 
\label{1}
\eea
where
\beq
s(t)=\sqrt{s_0^2+\frac{\eta}{\rho_r} t}.
\label{2}
\eeq

We note that 
\beq
\int_{-\infty}^\infty dx \delta(x,t) = \sqrt{2\pi} \delta_0 s_0 = {\rm const},
\eeq
which means that the linearized energy excess or deficit in the pulse does not change with time.  This 
can be seen directly by integrating Eq.~(\ref{conservation}) over $x$.  The density perturbation at the peak of the pulse is 
\beq
\delta(x=c_s t, t)= \delta_0 \frac{s_0}{\sqrt{s_0^2 + \frac{\eta}{\rho_r} t}}.
\eeq
This agrees with the qualitative expectations in Eqs.~(\ref{sM}), (\ref{deltas}) if we make the identification $\lambda_D^2 \sim (\eta/\rho_r)t \sim \tau_\gamma t$.

The total energy per unit area of the pulse is 
\beq
{\cal E} = \int d x (T^{00} -\rho_r) \sim {\cal E}_{0}+ {\cal E}_{s} ,
\label{Ew}
\eeq
where 
\beq
{\cal E}_{0} = \int dx \rho_r \delta = {\rm const} 
\eeq
and  
\beq
{\cal E}_{s} \sim \frac{1}{4}\int dx \rho_r\delta^2
\label{EMs}
\eeq
can be interpreted as the sound wave energy (per unit area).  This is the energy dissipated by Silk damping.

\subsection{$\mu$-distortion}
\label{sec:mu-planewave}

The $\mu$-distortion produced by the pulse should satisfy
\beq
{\dot\mu}(x,t)=-\frac{1}{4}\left(\frac{\partial}{\partial t}+c_s \frac{\partial}{\partial x}\right) \delta^2(x,t).
\label{3}
\eeq
The differential operator here is chosen so that ${\dot\mu}=0$ for a pulse without dissipation, $\delta(x-c_s t)$. Substituting (\ref{1}) in (\ref{3}), we have
\beq
\frac{\partial \mu}{\partial t}=\frac{\eta}{4\rho_r s^2(t)} \left[1-\frac{(x-c_s t)^2}{s^2(t)}\right] \delta^2(x,t).
\eeq

At each location $x$, $\delta(x,t)$ is significantly different from zero only for a period of time $\Delta t \sim s(t)/c_s \ll t$. During this period $s(t)$ changes very little and can be treated as a constant, 
\beq
s_x \equiv s(x/c_s) =\left(s_0^2 +\frac{\eta}{\rho_r}\frac{x}{c_s}\right)^{1/2}.  
\eeq
Then, after the pulse has passed the chemical potential acquires the value
\beq
\mu(x) \approx \frac{\eta \delta_0^2 s_0^2}{4\rho_r s_x^4}\int_{-\infty}^\infty dt \left(1-\frac{c_s^2 t^2}{s_x^2}\right) e^{-\frac{c_s^2 t^2}{s_x^2}}.
\eeq
Performing the Gaussian integrals, we obtain
\beq
\mu(x)=\frac{\sqrt{\pi}}{8} \frac{\eta\delta_0^2 s_0^2}{c_s \rho_r} \left(s_0^2+\frac{\eta}{\rho_r} \frac{x}{c_s}\right)^{-3/2} .
\eeq

We see that $\mu$ is generated everywhere where the pulse has passed.  As the pulse is dissipated, $\mu$ gets smaller at larger values of $x$.  If the $\mu$ era begins at $t=0$, when the pulse is at $x=0$, and then the photons mix up within a length $L\gg c_s s_0^2 /\tau_\gamma$, the resulting chemical potential would be
\beq
{\bar\mu} = \frac{1}{L}\int_0^L dx \mu(x) \sim \delta_0^2 \frac{s_0}{L} \sim \frac{{\cal E}_{0s}}{{\cal E}_r}.
\label{bar-mu}
\eeq
Here, ${\cal E}_{0s}\sim \rho_r \delta_0^2 s_0$ is the initial energy of sound waves per unit area of the pulse and ${\cal E}_r\sim \rho_r L$ is the total energy of radiation (also per unit area).

\bibliography{reference}

\end{document}